\documentclass[aps,prl,reprint,groupedaddress,longbibliography]{revtex4-1}
\usepackage{amsmath, braket}
\usepackage{pstricks}
\usepackage{graphicx} 

\DeclareMathOperator\arctanh{arctanh}

\usepackage{lipsum}

\newcommand\blfootnote[1]{%
  \begingroup
  \renewcommand\thefootnote{}\footnote{#1}%
  \addtocounter{footnote}{-1}%
  \endgroup
}
\makeatletter
\def\blfootnote{\gdef\@thefnmark{}\@footnotetext}
\makeatother
\begin{document}


\title{On the scalability of parametric down-conversion for generating higher-order Fock states}


\author{Johannes Tiedau$^{1}$}
\email{johannes.tiedau@uni-paderborn.com}
\author{Tim J. Bartley$^{1}$}
\author{Georg Harder$^{1}$}
\author{Adriana E. Lita$^{2}$}
\author{Sae Woo Nam$^{2}$}
\author{Thomas Gerrits$^{2}$}
\author{Christine Silberhorn$^{1}$}
\affiliation{$^{1}$Integrated Quantum Optics Group, Applied Physics, University of Paderborn, 33098 Paderborn, Germany}
\affiliation{$^{2}$National Institute of Standards and Technology, 325 Broadway, Boulder, CO 80305, USA}
\newcommand{\JT}[1]{\textcolor{red}{#1}}

\date{\today}

\begin{abstract}
Spontaneous parametric down-conversion (SPDC) is the most widely-used method to generate higher-order Fock states ($n\geq 2$). Yet, a consistent performance analysis from fundamental principles is missing. Here we address this problem by introducing a framework for state fidelity and generation probability under the consideration of losses and multimode emission. With this analysis we show the fundamental limitations of this process as well as a trade-off between state fidelity and generation rate intrinsic to the probabilistic nature of the process. This identifies the parameter space for which SPDC is useful when generating higher-order Fock states for quantum applications. We experimentally investigate the strong pump regime of SPDC and demonstrate heralded Fock states up to $\ket{n}=4$. 
\end{abstract}


\maketitle

\paragraph*{Introduction \ --}

As a representation of a discrete and well-defined number of excitations of the quantized electromagnetic fields, photon number states or Fock states are of significant fundamental and practical interest in quantum optics. They are the building blocks from which a range of exotic states may be constructed~\cite{Ourjoumtsev2007}, and find direct utility in metrology~\cite{Holland1993,Nagata2007,Slussarenko2017} and quantum information processing protocols~\cite{Yamamoto1986}.
Common to all these applications is that they become more advantageous with the size $n$ of the Fock state $\ket{n}$. However, generating higher-order Fock states becomes a challenging task and different approaches have been investigated to generate them~\cite{DAriano2000,Sanaka2005,Brown2003,McCusker2009,Glebov2014}.
To date, the most common process is strongly-pumped spontaneous parametric down-conversion (SPDC) in a nonlinear material~\cite{Ourjoumtsev2006, Waks2006,Zavatta2008,Cooper2013,Harder2016}. 
 
\begin{figure}[h!]
   \includegraphics[width=\linewidth]{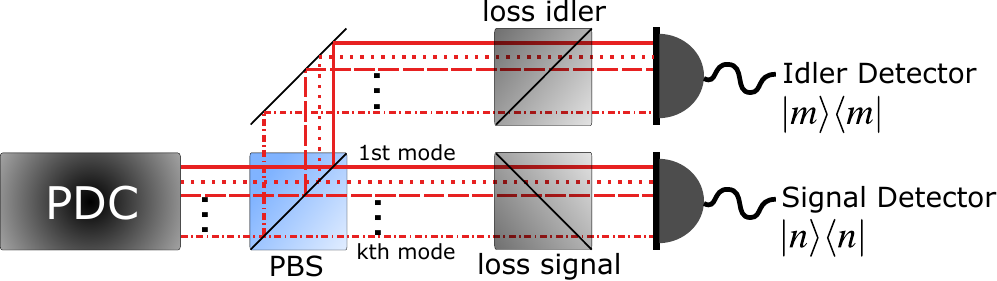}
\caption{Generating higher order Fock states with type~II parametric down-conversion (PDC). Signal and idler mode are split one a polarizing beam splitter (PBS). Losses as well as spectral multimodeness are considered. For further information see text.}
\label{fig:scheme}
\end{figure}

This consists of a nondeterministic decay of pump photons into precisely correlated numbers of photons in two modes (signal and idler). Increasing the number of pump photons increases the chance of multiple decays, resulting in signal and idler modes with higher occupation numbers.
In collinear Type~II or non-degenerate Type~I SPDC, these two modes are distinguishable (in polarization or frequency, respectively). The photon number correlations between the modes can be exploited to ``herald'' the generation of a particular Fock state: a projective measurement onto a specific photon number $n$ of the idler mode will result in the preparation of an $\ket{n}$ photon Fock state in the signal mode (see Fig.~\ref{fig:scheme}). 

Although heralded SPDC is the most widely used method to generate higher order Fock states, a detailed study of the optimal parameter range beyond the special cases for heralding one or two photons~\cite{Branczyk2010,Christ2012} is missing. 
In particular, the interplay of heralding photon numbers arising from different underlying spectral modes (Schmidt modes) renders this problem highly non-trivial.

In this letter we investigate the effects of losses and spectral multimodeness on heralding probability and fidelity of the produced state to an $n$ photon Fock state $\ket{n}$. This reveals the fundamental limits to generate higher-order Fock states with SPDC and shows a general trade-off between state fidelity and heralding probability. 
 
Furthermore we compare our findings with experimental results from a ppKTP waveguide source which has shown high brightness~\cite{Harder2016} and which can be engineered to emit light into a single mode~\cite{Eckstein2011}. This implementation investigates for the first time different pump intensities to maximize the generation probability of higher order Fock states. To simplify our analysis, we neglect higher order non-linear effects and time ordering~\cite{Christ2013, Quesada2015}.

\def\svgwidth{2in}

\paragraph*{Theoretical description \ --}
We model the state generated by a Type~II SPDC process with a 
general two-mode squeezed multi-spectral-mode PDC state given by 
\begin{equation}
\ket{\psi} = \bigotimes_k \sqrt{1-|\Lambda_k|^2} \sum_n \Lambda_k^n \ket{n,n}_k~,
\label{eq:PDC}
\end{equation}
where $k$ labels the modes, $\Lambda_k= \tanh(r_k)$ specifies the squeezing in dependence of the squeezing parameter $r$ and $n$ is the photon number. In the scope of this work we will consider different spectral modes $k$. The photon number probability $p_n$ for each spectral mode in Eq.~\ref{eq:PDC} is given by a geometric distribution. Without additional experimental effort it is not possible to distinguish the different spectral modes with standard photon detectors. 
Mathematically this is described by the convolution of the different spectral modes. Calculating the resulting distribution is a known problem and has been solved previously with generating functions \cite{Mauerer2009}. Here we develop a framework to describe the probability distribution by a discrete phase-type distribution \cite{Neuts1981}. It has the advantage of minimal computational effort as only matrix multiplications need to be carried out. In order to calculate the convoluted state distribution the vacuum probabilities $q_\text{k}=1-|\Lambda_k|^2$ for the different modes $k$ need to be written as a matrix of the following form
\begin{equation}
M = \begin{bmatrix}
    1-q_1 & q_1 & 0 & \dots  & 0 \\
   	0 & 1-q_2 & q_2 & \dots  & 0 \\
   	0 & 0 & 1-q_3 & \dots  & 0 \\
    \vdots & \vdots & \vdots & \ddots & \vdots \\
   0 & 0 & 0 & \dots  & 1-q_{K_\text{max}}
\end{bmatrix}~.
\label{eq:05}
\end{equation}
Then the probability of finding $n$ photons in up to $K_\text{max}$ modes with vacuum probabilities $q = (q_1,q_2,...,q_{K_\text{max}})$ is given by    
\begin{equation}
p_{n}(q) = \alpha M^{n+K_\text{max}-1} M_0~,
\label{eq:general_case}
\end{equation}
where $\alpha = (1,0,...,0)$ and $M_0 = (0,0,...,q_{K_\text{max}})^T$. 
$K_\text{max}$ defines the number of modes that are considered. An even simpler solution is possible if all spectral modes have the same vacuum probability (see supplement). 
From now on, we will assume Gaussian functions for the pump spectrum and for the phase matching, which is a valid assumption if, for example,  spectral filtering or apodized poling \cite{Branczyk2011,Graffitti2017} is used. In this case, the squeezing parameter $r_k$ for the kth mode is exponentially decreasing \cite{URen2003}
\begin{equation}
\begin{aligned}
	r_k &= B \cdot \lambda_k  \\
	\lambda_k &= \sqrt{(1-\mu^2)}\mu^{k-1}~,
\label{eq:optical_gain}
\end{aligned}
\end{equation}
where $B$ is the optical gain that defines the squeezing strength and $\mu \in [0,1)$ determines the effective number of spectral modes known as the Schmidt number $K= 1/\sum_k(\lambda_k^4)$. We want to stress here that this definition of the Schmidt number does not depend on the optical gain. Any measure that depends on the absolute number of photons per mode however will change (for example the $g^{(2)}(0)$ correlation function, for further details see supplemental material).

Beside the effects arising from multiple modes, we will also consider losses in the detection process. Losses can be considered by an appropriate measurement POVM \cite{Silberhorn2007}. Additional effects for example from dark counts or from time multiplexed detection schemes are not considered here but can be added to the model  \cite{Sperling2012}. 

We will consider the fidelity towards the desired Fock-state in a predefined spectral mode (dominant Schmidt mode) as a measure for the quality of the heralded state \cite{Jozsa1994}. As used in \cite{Christ2012} we will use the heralding probability and the fidelity as described above as the benchmark parameters.

With this model it is now possible to investigate the multimode heralding process. At first we will consider the effect of losses in the heralding (idler) arm for a single spectral mode. Losses in the signal arm will be considered in the experimental section to match the theoretical model to the measured data.  

\begin{figure}[h!]
   \includegraphics[width=\linewidth]{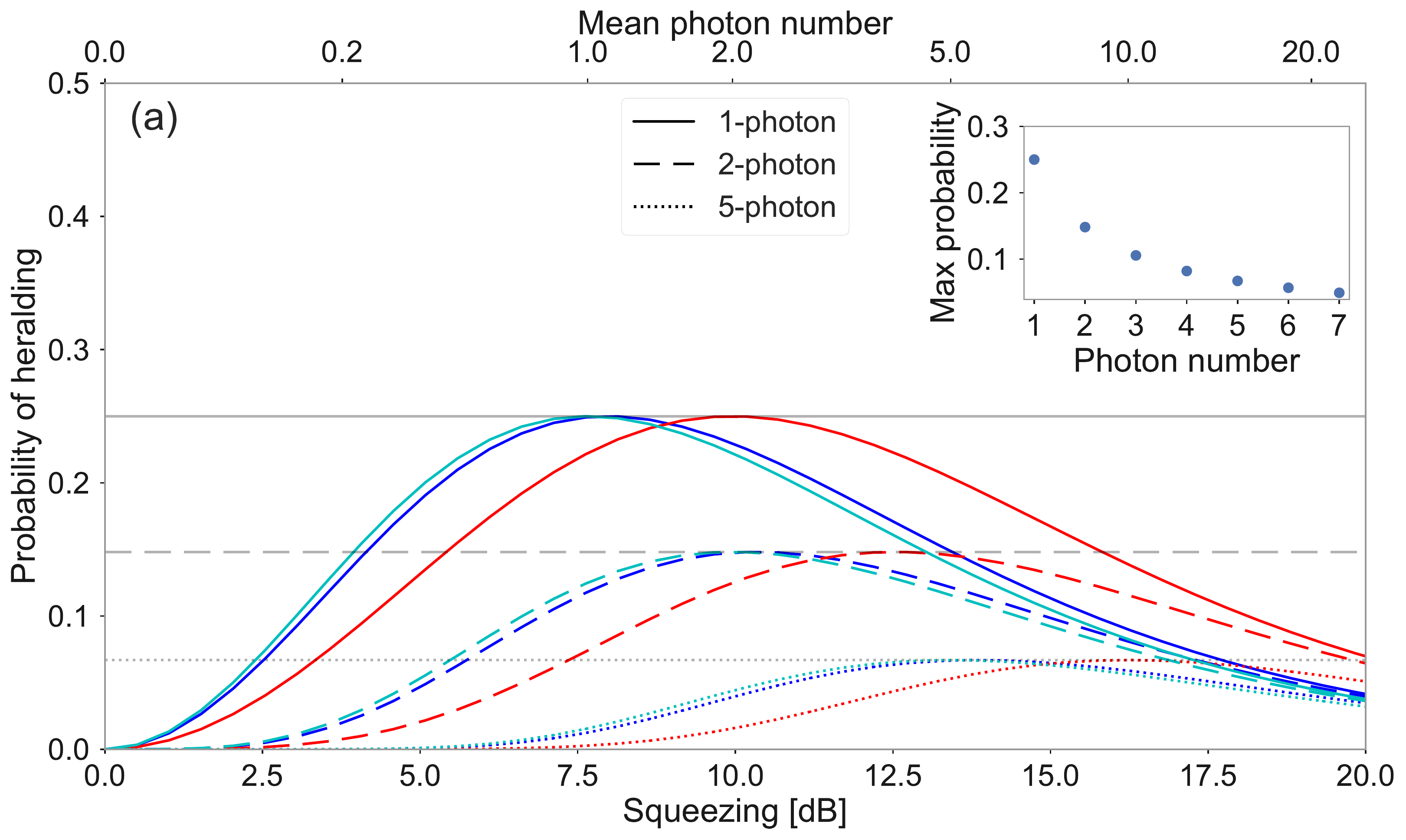}
   \includegraphics[width=\linewidth]{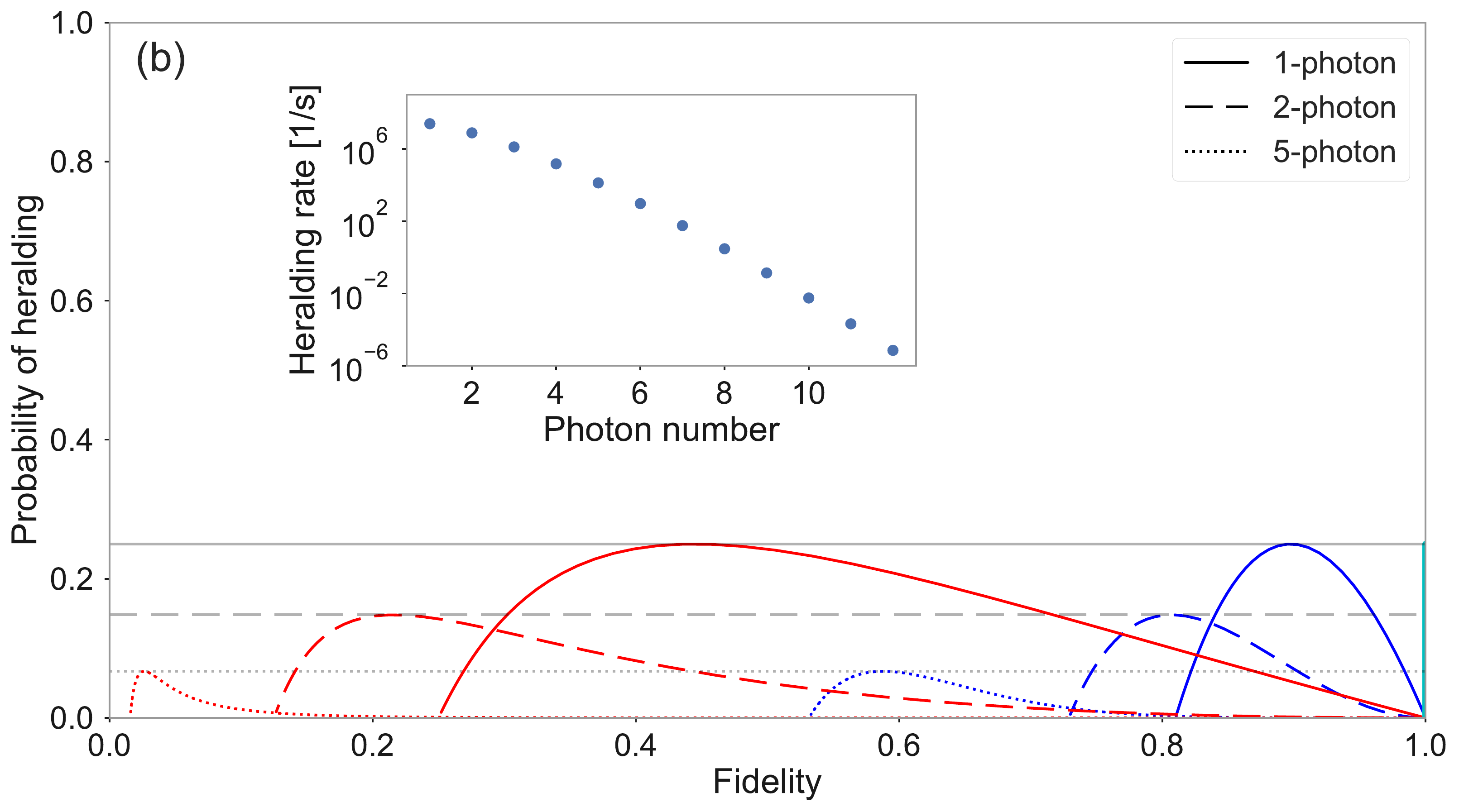}
\caption{Heralding probability and state fidelity are calculated for the heralding transmission values of 1.0 (cyan), 0.9 (blue) and 0.5 (red) and a heralded single photon, 2-photon and 5-photon state. (a): Dependence of heralding probability on the initial squeezing parameter. The inset shows the maximal generation probability of an n-photon state for a single mode source. This value is also indicated by grey horizontal lines in both figures. (b):  Relation of heralding probability and state fidelity while the squeezing value is changed. For a heralding transmission value of 1.0 (cyan) unit fidelity is always reached. The bottom inset shows the generation rate for Fock states under optimistic experimental assumptions (see main text). }
\label{fig:loss}
\end{figure}

The effect of losses in the heralding arm (idler) is shown in Fig.~\ref{fig:loss}. Heralding probability and state fidelity are plotted versus the initial squeezing value for a single mode state. The blue curve assumes a heralding efficiency of 90\%, whereas the red curve is calculated for a heralding efficiency of 50\%. It can be seen that losses in the idler arm are not critical for the heralding probability because the photon-number distribution of the heralding mode (heralded mode is traced out) is a thermal state and thermal states stay thermal under losses. In principle, a lower transmission value can always be counteracted by increasing the squeezing parameter i.e. increasing the pump power of the SPDC process. The geometric distribution of the thermal state directly reveals the maximal heralding probability for an $n$ - photon Fock state generated by a single mode source (cf. top inset Fig.~\ref{fig:loss}) of
\begin{equation}
p_\text{max,K=1}(n) = \frac{n^n}{(1+n)^{1+n}}~.
\label{eq:maxPsingle}
\end{equation} 
The fidelity, on the other hand, is obviously influenced by losses as photon number correlations between signal and idler mode are affected. The effect of loss in the heralding arm can be decreased by reducing the squeezing value; unit fidelity can always be reached in principle for vanishing generation probability. 
Fig~\ref{fig:loss} (b) illustrates the fundamental limits of SPDC for higher-order Fock state generation as well as the trade-off between generation probability and state fidelity (the heralding probability is limited and cannot reach the maximum simultaneously with the fidelity). This effect becomes more pronounced for higher photon numbers and for higher losses in the heralding arm. This is further demonstrated in the inset to Fig~\ref{fig:loss} (b), which shows the generation rate for Fock states under optimistic experimental assumptions in term of experimental repetition rate (100~MHz), heralding efficiency (90\%) and target fidelity (90\%). With these restrictions it can be shown that only Fock states up to n=9 can be realized at 0.1 event/s. Further details about realistic limits in experiments are provided in the supplemental material.

\begin{figure}[h!]
   \includegraphics[width=\linewidth]{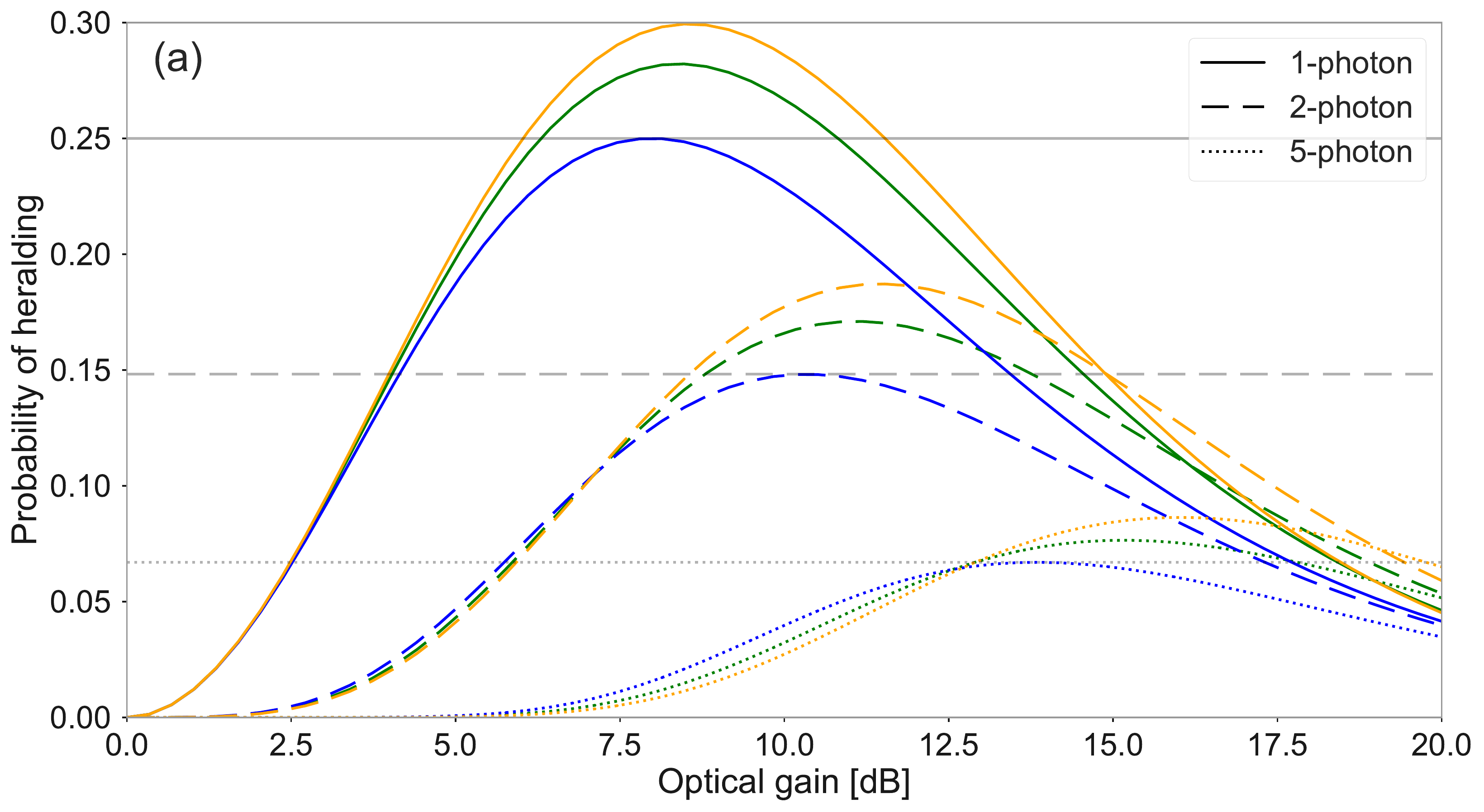}
   \includegraphics[width=\linewidth]{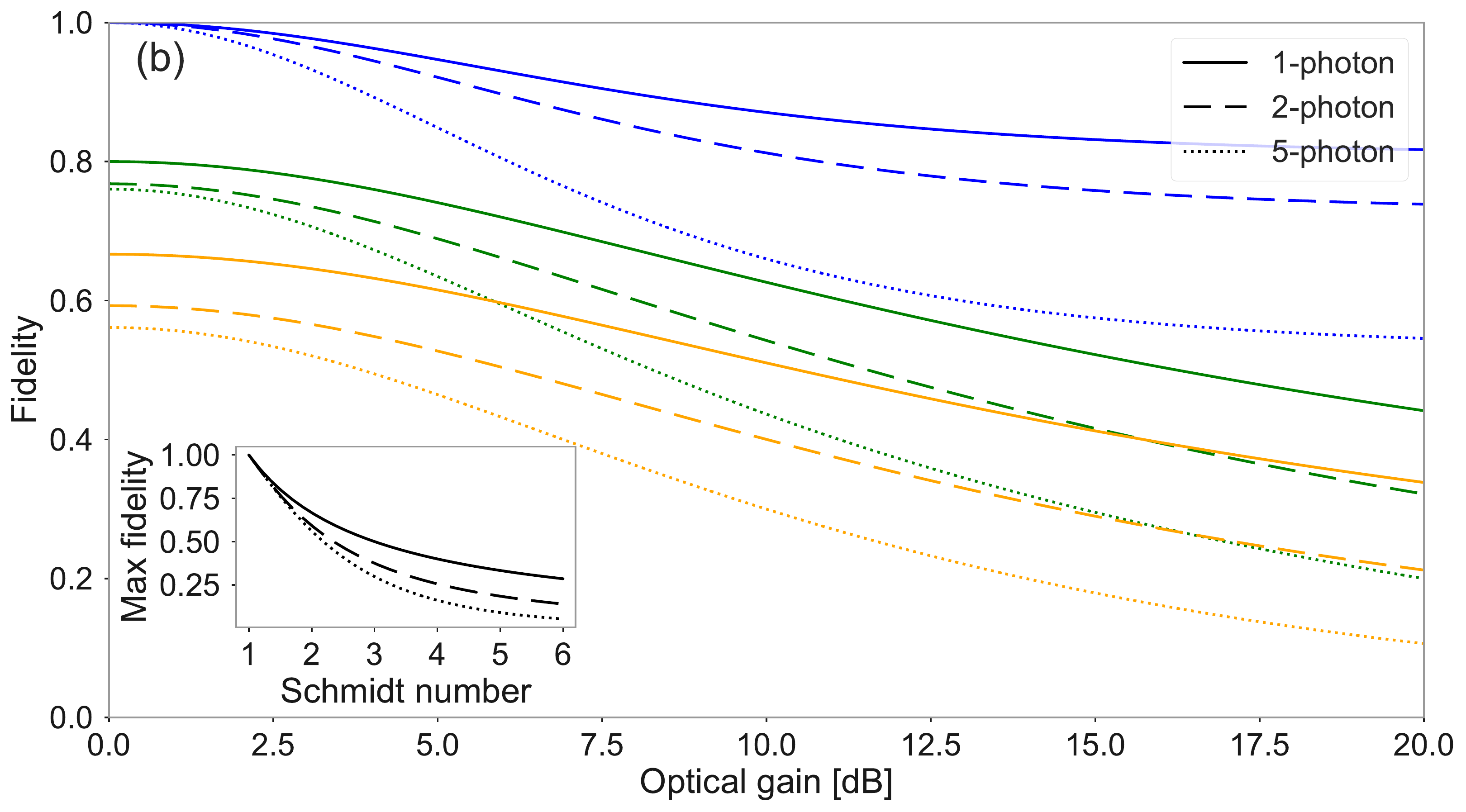}
\caption{Heralding probability (a) and state fidelity (b) versus the optical gain parameter. For these plots the heralding transmission value is kept constant at 0.9. Three different Schmidt values of 1, 1.5 and 2 are plotted (blue, green and yellow curve respectively).  The inset (b) shows how the Schmidt number influences the maximal fidelity. Spectral modes up to $K_\text{max} = 35$ were considered for these calculations.}
\label{fig:mode}
\end{figure}

Secondly, we can investigate effects from multiple spectral modes as illustrated in Fig.~\ref{fig:mode}. Here the generation probability and state fidelity are shown versus the optical gain, which is used as the generalization of the squeezing parameter for the multimode case (cf. Eq.~\ref{eq:optical_gain}). The maximal generation probability increases with more modes present as the mode distribution goes from being a thermal state to a Possonian distribution $p_\text{max}(n) = e^{-n}n^n/n!$. This is unsuitable for many applications since it will decrease the spectral purity of the heralded state. This relation of the Schmidt number and the maximal achievable fidelity is shown in the inset Fig.~\ref{fig:mode}. Despite this limit, the maximal probability can be used to determine the Schmidt number of the generated state. In general, measuring the Schmidt number in the high gain regime is a difficult task. For example the $g^2(0)$ value, which is typically used to determine the mode number, is not directly related to the Schmidt number in this regime \cite{Christ2011}. However, if the maximal heralding probability is used to determine the Schmidt number it should be noted that losses can change the probability distribution of convoluted thermal states and therefore can change the maximal heralding probability. This result is discussed in detail in the supplemental material.

\paragraph*{Experimental setup and results\ --}

\begin{figure}[h]
\includegraphics[width=\linewidth]{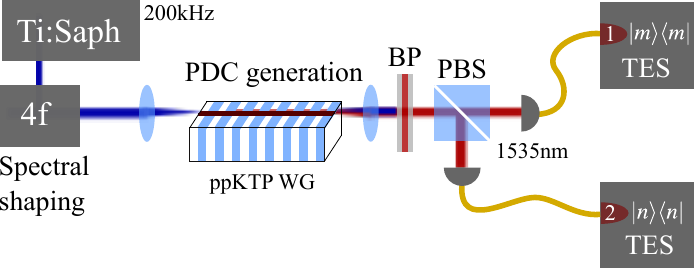}
\caption{Experimental setup as used in \cite{Harder2016}. Pulsed light from a Ti:Sapphire laser is spectrally filtered by a 4f line and coupled into a periodically poled KTP waveguide. A bandpass (BP) filter is used to filter out the pump as well as to suppress sinc-sidelobes of the phase matching function. Signal and idler are split on a polarizing beam splitter, (PBS) coupled into fibers and detected by intrinsic photon number resolving transition edge sensors (TES).}
\label{fig:exp_setup}
\end{figure}

To confirm our theoretical findings, we experimentally generated higher order Fock states and measured their heralding probabilities and fidelities. For this we used a type-II parametric down-conversion process (PDC) in a periodically poled potassium titanyl phosphate (KTP) waveguide. This crystal is pumped with pulsed light from a Ti:Sapphire oscillator at 767.5 nm (see Fig.~\ref{fig:exp_setup}). The source is used for its highly single mode performance \cite{Eckstein2011} and extremely high brightness  \cite{Harder2016}. The spectral purity of this source depends on the phase matching and pump parameters \cite{Grice2001}. Here we use a non-optimal pump bandwidth in order to see an increased effect of multiple spectral modes \cite{Avenhaus2008}. Detection is performed with intrinsic photon number resolving transition edge sensors \cite{Lita2008} (TES). These detectors offer near unit efficiency and extremely high photon number discrimination in the few photon regime \cite{Humphreys2015}.
Details about the conversion from TES response functions to photon numbers can be found in the supplementary material. 

\begin{figure}[h!]
\includegraphics[width=\linewidth]{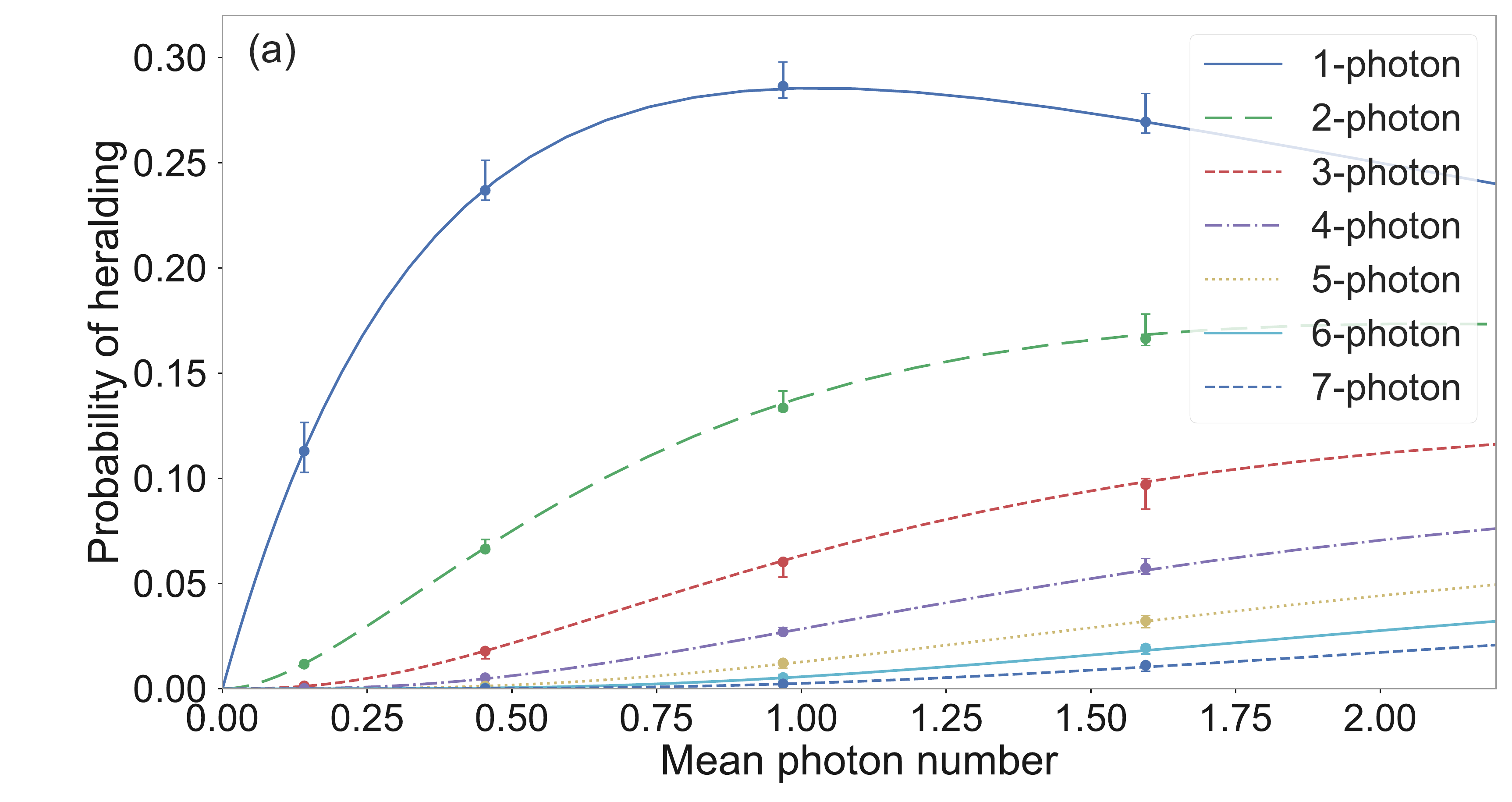}
\includegraphics[width=\linewidth]{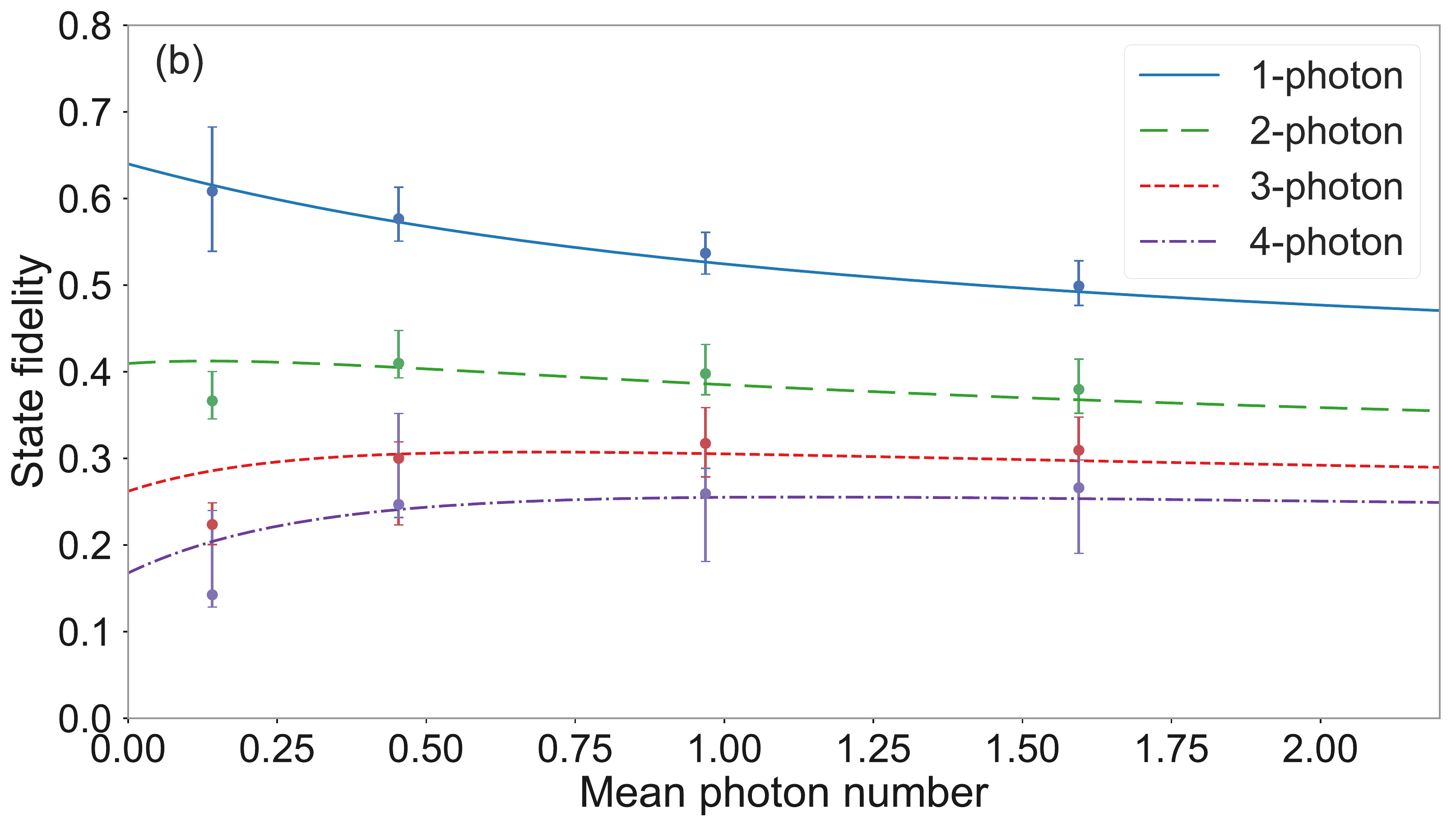}
\caption{Experimental data (points) for the heralding probability (a) and the state fidelity without correcting for losses (b) versus the measured mean photon number. Four different pump intensities were investigated. Colored lines show the theoretical values with $K=1.61$, $\eta_i=0.59$ and $\eta_s=0.64$ as the only free parameters. 
In (b) the state fidelity to a spectrally multimode state is plotted. The state fidelity to a single-mode state is shown in the supplement. See text for further details.}
\label{fig:exp}
\end{figure}

We analyzed the heralding probability and state fidelity for four different pump intensities. The results are shown in Fig.~\ref{fig:exp}. Heralding probabilities are analyzed for up to 7 photons. State fidelities are evaluated for up to 4 photons on the heralding detector. Beyond four photons, the statistical uncertainties as well as errors arising from photon number identification (cf. \cite{Humphreys2015}) become substantial. 
Colored lines show fitted curves based on the theory shown above. Only one set of fitting parameters was used for all curves (Schmidt number $K = 1.61$ and $\eta_i = 0.59$ and $\eta_s = 0.64$ for the transmission of the idler and signal arm respectively). The experimental data can be described with very high precision with our theory. Error bars are discussed in the supplement. Small deviations can be most likely explained by coupling drifts in the setup between the four measurement runs.  

The setup shown in Fig.~\ref{fig:exp_setup} is not able to distinguish spectral modes to determine the single-mode state fidelity directly. Instead the convolution of all modes was measured and compared to theory with $K=1.61$ (compare Fig.~\ref{fig:exp}). The corollary of this is that this measurment can be used to extract the modal behaviour even without a mode resolving measurement, by fitting the experimental data with our multimode theoretical model. A similar approach has also been shown in \cite{Burenkov2017}.

It can be seen that the measured heralding probability for different heralded Fock states is higher than the single mode theory predicts (e.g. the 1-photon curve in Fig.~\ref{fig:exp}(a) goes above 25\%). This shows the importance of considering a multimode theory even though fewer than two Schmidt modes are present. The theory in this paper excludes time ordering but when searching for time ordering phenomena the effects presented here should be considered.
Additionally it can be seen that rather high mean photon numbers are required to herald higher order Fock states with reasonable probability. In order to generate these squeezing values in a single pass configuration, waveguided nonlinear materials are essential.  
If losses for the signal (heralded photon) are included (as shown in Fig.~\ref{fig:exp}) more optical gain can be beneficial to increase the state fidelity. In this case losses and higher order contributions increase the generation probability of the desired state. However, for low signal losses, which are required for most applications were higher order Fock states are involved, the state fidelity is strictly decreasing for higher optical gain.

\paragraph*{Conclusion \ --}

Spontaneous parametric down-conversion is the most widely used tool to generate higher-order optical Fock states. In this letter the limitations of this process in terms of generation probability and state fidelity under the consideration of losses and spectral multimodeness were investigated. We found absolute limits for the generation of Fock states in SPDC as well as a fundamental trade-off between generation probability and state fidelity for low signal losses. This means that high fidelity Fock states under realistic experimental constraints 
and generation probabilities cannot be realized beyond $n=9$ at 0.1~event/s. 
We have experimentally investigated different pump intensities to maximize the generation probability of higher-order Fock states and showed that waveguided spectrally engineered sources offer many essential advantages for pulsed SPDC as they allow for high generation probabilities and single mode emission. With this result is it possible to calculate the feasibility of new experiments requiring higher order Fock states. At the same time this stresses that alternative approaches to generate higher-order Fock states, as for example shown in \cite{McCusker2009,Glebov2014}, need be explored further.

\begin{acknowledgments}
We thank Evan Meyer-Scott, Ivan Burenkov and Robinjeet Singh for helpful discussions during the preparation of the manuscript.  This work was supported by the European Union's Quantum Flagship research and innovation initiative under project agreement (820365) and by the Quantum Information Science Initiative (QISI). TB acknowledges funding by Deutsche Forschungsgemeinschaft (DFG, German Research Foundation)
– Projektnummer 231447078 – TRR 142. Contribution of NIST, an agency of the U.S. government, not subject to copyright.
\end{acknowledgments}

\bibliography{MyCollection}

\begin{thebibliography}{37}%
\makeatletter
\providecommand \@ifxundefined [1]{%
 \@ifx{#1\undefined}
}%
\providecommand \@ifnum [1]{%
 \ifnum #1\expandafter \@firstoftwo
 \else \expandafter \@secondoftwo
 \fi
}%
\providecommand \@ifx [1]{%
 \ifx #1\expandafter \@firstoftwo
 \else \expandafter \@secondoftwo
 \fi
}%
\providecommand \natexlab [1]{#1}%
\providecommand \enquote  [1]{``#1''}%
\providecommand \bibnamefont  [1]{#1}%
\providecommand \bibfnamefont [1]{#1}%
\providecommand \citenamefont [1]{#1}%
\providecommand \href@noop [0]{\@secondoftwo}%
\providecommand \href [0]{\begingroup \@sanitize@url \@href}%
\providecommand \@href[1]{\@@startlink{#1}\@@href}%
\providecommand \@@href[1]{\endgroup#1\@@endlink}%
\providecommand \@sanitize@url [0]{\catcode `\\12\catcode `\$12\catcode
  `\&12\catcode `\#12\catcode `\^12\catcode `\_12\catcode `\%12\relax}%
\providecommand \@@startlink[1]{}%
\providecommand \@@endlink[0]{}%
\providecommand \url  [0]{\begingroup\@sanitize@url \@url }%
\providecommand \@url [1]{\endgroup\@href {#1}{\urlprefix }}%
\providecommand \urlprefix  [0]{URL }%
\providecommand \Eprint [0]{\href }%
\providecommand \doibase [0]{http://dx.doi.org/}%
\providecommand \selectlanguage [0]{\@gobble}%
\providecommand \bibinfo  [0]{\@secondoftwo}%
\providecommand \bibfield  [0]{\@secondoftwo}%
\providecommand \translation [1]{[#1]}%
\providecommand \BibitemOpen [0]{}%
\providecommand \bibitemStop [0]{}%
\providecommand \bibitemNoStop [0]{.\EOS\space}%
\providecommand \EOS [0]{\spacefactor3000\relax}%
\providecommand \BibitemShut  [1]{\csname bibitem#1\endcsname}%
\let\auto@bib@innerbib\@empty
\bibitem [{\citenamefont {Ourjoumtsev}\ \emph {et~al.}(2007)\citenamefont
  {Ourjoumtsev}, \citenamefont {Jeong}, \citenamefont {Tualle-Brouri},\ and\
  \citenamefont {Grangier}}]{Ourjoumtsev2007}%
  \BibitemOpen
  \bibfield  {author} {\bibinfo {author} {\bibfnamefont {Alexei}\ \bibnamefont
  {Ourjoumtsev}}, \bibinfo {author} {\bibfnamefont {Hyunseok}\ \bibnamefont
  {Jeong}}, \bibinfo {author} {\bibfnamefont {Rosa}\ \bibnamefont
  {Tualle-Brouri}}, \ and\ \bibinfo {author} {\bibfnamefont {Philippe}\
  \bibnamefont {Grangier}},\ }\bibfield  {title} {\enquote {\bibinfo {title}
  {{Generation of optical 'Schr{\"{o}}dinger cats' from photon number
  states}},}\ }\href {\doibase 10.1038/nature06054} {\bibfield  {journal}
  {\bibinfo  {journal} {Nature}\ }\textbf {\bibinfo {volume} {448}},\ \bibinfo
  {pages} {784--786} (\bibinfo {year} {2007})}\BibitemShut {NoStop}%
\bibitem [{\citenamefont {Holland}\ and\ \citenamefont
  {Burnett}(1993)}]{Holland1993}%
  \BibitemOpen
  \bibfield  {author} {\bibinfo {author} {\bibfnamefont {M~J}\ \bibnamefont
  {Holland}}\ and\ \bibinfo {author} {\bibfnamefont {K}~\bibnamefont
  {Burnett}},\ }\bibfield  {title} {\enquote {\bibinfo {title}
  {{Interferometric Detection of Optical Phase Shifts at t h e Heisenberg
  Limit}},}\ }\href@noop {} {\bibfield  {journal} {\bibinfo  {journal}
  {Physical Review Letters}\ }\textbf {\bibinfo {volume} {71}},\ \bibinfo
  {pages} {1355--1358} (\bibinfo {year} {1993})}\BibitemShut {NoStop}%
\bibitem [{\citenamefont {Nagata}\ \emph {et~al.}(2007)\citenamefont {Nagata},
  \citenamefont {Okamoto}, \citenamefont {O'Brien}, \citenamefont {Sasaki},\
  and\ \citenamefont {Takeuchi}}]{Nagata2007}%
  \BibitemOpen
  \bibfield  {author} {\bibinfo {author} {\bibfnamefont {Tomohisa}\
  \bibnamefont {Nagata}}, \bibinfo {author} {\bibfnamefont {Ryo}\ \bibnamefont
  {Okamoto}}, \bibinfo {author} {\bibfnamefont {Jeremy~L.}\ \bibnamefont
  {O'Brien}}, \bibinfo {author} {\bibfnamefont {Keiji}\ \bibnamefont {Sasaki}},
  \ and\ \bibinfo {author} {\bibfnamefont {Shigeki}\ \bibnamefont {Takeuchi}},\
  }\bibfield  {title} {\enquote {\bibinfo {title} {{Beating the standard
  quantum limit with four-entangled photons}},}\ }\href {\doibase
  10.1126/science.1138007} {\bibfield  {journal} {\bibinfo  {journal}
  {Science}\ }\textbf {\bibinfo {volume} {316}},\ \bibinfo {pages} {726--729}
  (\bibinfo {year} {2007})}\BibitemShut {NoStop}%
\bibitem [{\citenamefont {Slussarenko}\ \emph {et~al.}(2017)\citenamefont
  {Slussarenko}, \citenamefont {Weston}, \citenamefont {Chrzanowski},
  \citenamefont {Shalm}, \citenamefont {Verma}, \citenamefont {Nam},\ and\
  \citenamefont {Pryde}}]{Slussarenko2017}%
  \BibitemOpen
  \bibfield  {author} {\bibinfo {author} {\bibfnamefont {Sergei}\ \bibnamefont
  {Slussarenko}}, \bibinfo {author} {\bibfnamefont {Morgan~M.}\ \bibnamefont
  {Weston}}, \bibinfo {author} {\bibfnamefont {Helen~M.}\ \bibnamefont
  {Chrzanowski}}, \bibinfo {author} {\bibfnamefont {Lynden~K.}\ \bibnamefont
  {Shalm}}, \bibinfo {author} {\bibfnamefont {Varun~B.}\ \bibnamefont {Verma}},
  \bibinfo {author} {\bibfnamefont {Sae~Woo}\ \bibnamefont {Nam}}, \ and\
  \bibinfo {author} {\bibfnamefont {Geoff~J.}\ \bibnamefont {Pryde}},\
  }\bibfield  {title} {\enquote {\bibinfo {title} {{Unconditional violation of
  the shot-noise limit in photonic quantum metrology}},}\ }\href {\doibase
  10.1038/s41566-017-0011-5} {\bibfield  {journal} {\bibinfo  {journal} {Nature
  Photonics}\ }\textbf {\bibinfo {volume} {11}},\ \bibinfo {pages} {700--703}
  (\bibinfo {year} {2017})}\BibitemShut {NoStop}%
\bibitem [{\citenamefont {Yamamoto}\ and\ \citenamefont
  {Haus}(1986)}]{Yamamoto1986}%
  \BibitemOpen
  \bibfield  {author} {\bibinfo {author} {\bibfnamefont {Y.}~\bibnamefont
  {Yamamoto}}\ and\ \bibinfo {author} {\bibfnamefont {H.~A.}\ \bibnamefont
  {Haus}},\ }\bibfield  {title} {\enquote {\bibinfo {title} {{Preparation,
  measurement and information capacity of optical quantum states}},}\ }\href
  {\doibase 10.1103/RevModPhys.58.1001} {\bibfield  {journal} {\bibinfo
  {journal} {Reviews of Modern Physics}\ }\textbf {\bibinfo {volume} {58}},\
  \bibinfo {pages} {1001--1020} (\bibinfo {year} {1986})}\BibitemShut {NoStop}%
\bibitem [{\citenamefont {D'Ariano}\ \emph {et~al.}(2000)\citenamefont
  {D'Ariano}, \citenamefont {Maccone}, \citenamefont {Paris},\ and\
  \citenamefont {Sacchi}}]{DAriano2000}%
  \BibitemOpen
  \bibfield  {author} {\bibinfo {author} {\bibfnamefont {G.~M.}\ \bibnamefont
  {D'Ariano}}, \bibinfo {author} {\bibfnamefont {L.}~\bibnamefont {Maccone}},
  \bibinfo {author} {\bibfnamefont {M.~G.A.}\ \bibnamefont {Paris}}, \ and\
  \bibinfo {author} {\bibfnamefont {M.~F.}\ \bibnamefont {Sacchi}},\ }\bibfield
   {title} {\enquote {\bibinfo {title} {{Optical Fock-state synthesizer}},}\
  }\href {\doibase 10.1103/PhysRevA.61.053817} {\bibfield  {journal} {\bibinfo
  {journal} {Physical Review A - Atomic, Molecular, and Optical Physics}\
  }\textbf {\bibinfo {volume} {61}},\ \bibinfo {pages} {5} (\bibinfo {year}
  {2000})}\BibitemShut {NoStop}%
\bibitem [{\citenamefont {Sanaka}(2005)}]{Sanaka2005}%
  \BibitemOpen
  \bibfield  {author} {\bibinfo {author} {\bibfnamefont {Kaoru}\ \bibnamefont
  {Sanaka}},\ }\bibfield  {title} {\enquote {\bibinfo {title} {{Linear optical
  extraction of photon-number Fock states from coherent states}},}\ }\href
  {\doibase 10.1103/PhysRevA.71.021801} {\bibfield  {journal} {\bibinfo
  {journal} {Physical Review A - Atomic, Molecular, and Optical Physics}\
  }\textbf {\bibinfo {volume} {71}},\ \bibinfo {pages} {1--4} (\bibinfo {year}
  {2005})}\BibitemShut {NoStop}%
\bibitem [{\citenamefont {Brown}\ \emph {et~al.}(2003)\citenamefont {Brown},
  \citenamefont {Dani}, \citenamefont {Stamper-Kurn},\ and\ \citenamefont
  {Whaley}}]{Brown2003}%
  \BibitemOpen
  \bibfield  {author} {\bibinfo {author} {\bibfnamefont {K.~R.}\ \bibnamefont
  {Brown}}, \bibinfo {author} {\bibfnamefont {K.~M.}\ \bibnamefont {Dani}},
  \bibinfo {author} {\bibfnamefont {D.~M.}\ \bibnamefont {Stamper-Kurn}}, \
  and\ \bibinfo {author} {\bibfnamefont {K.~B.}\ \bibnamefont {Whaley}},\
  }\bibfield  {title} {\enquote {\bibinfo {title} {{Deterministic optical
  Fock-state generation}},}\ }\href {\doibase 10.1103/PhysRevA.67.043818}
  {\bibfield  {journal} {\bibinfo  {journal} {Physical Review A - Atomic,
  Molecular, and Optical Physics}\ }\textbf {\bibinfo {volume} {67}},\ \bibinfo
  {pages} {16} (\bibinfo {year} {2003})}\BibitemShut {NoStop}%
\bibitem [{\citenamefont {McCusker}\ and\ \citenamefont
  {Kwiat}(2009)}]{McCusker2009}%
  \BibitemOpen
  \bibfield  {author} {\bibinfo {author} {\bibfnamefont {Kevin~T.}\
  \bibnamefont {McCusker}}\ and\ \bibinfo {author} {\bibfnamefont {Paul~G.}\
  \bibnamefont {Kwiat}},\ }\bibfield  {title} {\enquote {\bibinfo {title}
  {{Efficient optical quantum state engineering}},}\ }\href {\doibase
  10.1103/PhysRevLett.103.163602} {\bibfield  {journal} {\bibinfo  {journal}
  {Physical Review Letters}\ }\textbf {\bibinfo {volume} {103}},\ \bibinfo
  {pages} {1--4} (\bibinfo {year} {2009})}\BibitemShut {NoStop}%
\bibitem [{\citenamefont {Glebov}\ \emph {et~al.}(2014)\citenamefont {Glebov},
  \citenamefont {Fan},\ and\ \citenamefont {Migdall}}]{Glebov2014}%
  \BibitemOpen
  \bibfield  {author} {\bibinfo {author} {\bibfnamefont {Boris~L.}\
  \bibnamefont {Glebov}}, \bibinfo {author} {\bibfnamefont {Jingyun}\
  \bibnamefont {Fan}}, \ and\ \bibinfo {author} {\bibfnamefont
  {A.}~\bibnamefont {Migdall}},\ }\bibfield  {title} {\enquote {\bibinfo
  {title} {{Photon number squeezing in repeated parametric downconversion with
  ancillary photon-number measurements}},}\ }\href {\doibase
  10.1364/OE.22.020358} {\bibfield  {journal} {\bibinfo  {journal} {Optics
  Express}\ }\textbf {\bibinfo {volume} {22}},\ \bibinfo {pages} {20358}
  (\bibinfo {year} {2014})}\BibitemShut {NoStop}%
\bibitem [{\citenamefont {Ourjoumtsev}\ \emph {et~al.}(2006)\citenamefont
  {Ourjoumtsev}, \citenamefont {Tualle-Brouri},\ and\ \citenamefont
  {Grangier}}]{Ourjoumtsev2006}%
  \BibitemOpen
  \bibfield  {author} {\bibinfo {author} {\bibfnamefont {Alexei}\ \bibnamefont
  {Ourjoumtsev}}, \bibinfo {author} {\bibfnamefont {Rosa}\ \bibnamefont
  {Tualle-Brouri}}, \ and\ \bibinfo {author} {\bibfnamefont {Philippe}\
  \bibnamefont {Grangier}},\ }\bibfield  {title} {\enquote {\bibinfo {title}
  {{Quantum homodyne tomography of a two-photon fock state}},}\ }\href
  {\doibase 10.1103/PhysRevLett.96.213601} {\bibfield  {journal} {\bibinfo
  {journal} {Physical Review Letters}\ }\textbf {\bibinfo {volume} {96}},\
  \bibinfo {pages} {1--4} (\bibinfo {year} {2006})}\BibitemShut {NoStop}%
\bibitem [{\citenamefont {Waks}\ \emph {et~al.}(2006)\citenamefont {Waks},
  \citenamefont {Diamanti},\ and\ \citenamefont {Yamamoto}}]{Waks2006}%
  \BibitemOpen
  \bibfield  {author} {\bibinfo {author} {\bibfnamefont {Edo}\ \bibnamefont
  {Waks}}, \bibinfo {author} {\bibfnamefont {Eleni}\ \bibnamefont {Diamanti}},
  \ and\ \bibinfo {author} {\bibfnamefont {Yoshihisa}\ \bibnamefont
  {Yamamoto}},\ }\bibfield  {title} {\enquote {\bibinfo {title} {{Generation of
  photon number states}},}\ }\href {\doibase 10.1088/1367-2630/8/1/004}
  {\bibfield  {journal} {\bibinfo  {journal} {New Journal of Physics}\ }\textbf
  {\bibinfo {volume} {8}} (\bibinfo {year} {2006}),\
  10.1088/1367-2630/8/1/004}\BibitemShut {NoStop}%
\bibitem [{\citenamefont {Zavatta}\ \emph {et~al.}(2008)\citenamefont
  {Zavatta}, \citenamefont {Parigi},\ and\ \citenamefont
  {Bellini}}]{Zavatta2008}%
  \BibitemOpen
  \bibfield  {author} {\bibinfo {author} {\bibfnamefont {Alessandro}\
  \bibnamefont {Zavatta}}, \bibinfo {author} {\bibfnamefont {Valentina}\
  \bibnamefont {Parigi}}, \ and\ \bibinfo {author} {\bibfnamefont {Marco}\
  \bibnamefont {Bellini}},\ }\bibfield  {title} {\enquote {\bibinfo {title}
  {{Toward quantum frequency combs: Boosting the generation of highly
  nonclassical light states by cavity-enhanced parametric down-conversion at
  high repetition rates}},}\ }\href {\doibase 10.1103/PhysRevA.78.033809}
  {\bibfield  {journal} {\bibinfo  {journal} {Physical Review A - Atomic,
  Molecular, and Optical Physics}\ }\textbf {\bibinfo {volume} {78}},\ \bibinfo
  {pages} {1--4} (\bibinfo {year} {2008})}\BibitemShut {NoStop}%
\bibitem [{\citenamefont {Cooper}\ \emph {et~al.}(2013)\citenamefont {Cooper},
  \citenamefont {S{\"{o}}ller},\ and\ \citenamefont {Smith}}]{Cooper2013}%
  \BibitemOpen
  \bibfield  {author} {\bibinfo {author} {\bibfnamefont {Merlin}\ \bibnamefont
  {Cooper}}, \bibinfo {author} {\bibfnamefont {Christoph}\ \bibnamefont
  {S{\"{o}}ller}}, \ and\ \bibinfo {author} {\bibfnamefont {Brian~J.}\
  \bibnamefont {Smith}},\ }\bibfield  {title} {\enquote {\bibinfo {title}
  {{High-stability time-domain balanced homodyne detector for ultrafast optical
  pulse applications}},}\ }\href {\doibase 10.1080/09500340.2013.797612}
  {\bibfield  {journal} {\bibinfo  {journal} {Journal of Modern Optics}\
  }\textbf {\bibinfo {volume} {60}},\ \bibinfo {pages} {611--616} (\bibinfo
  {year} {2013})}\BibitemShut {NoStop}%
\bibitem [{\citenamefont {Harder}\ \emph {et~al.}(2016)\citenamefont {Harder},
  \citenamefont {Bartley}, \citenamefont {Lita}, \citenamefont {Nam},
  \citenamefont {Gerrits},\ and\ \citenamefont {Silberhorn}}]{Harder2016}%
  \BibitemOpen
  \bibfield  {author} {\bibinfo {author} {\bibfnamefont {Georg}\ \bibnamefont
  {Harder}}, \bibinfo {author} {\bibfnamefont {Tim~J.}\ \bibnamefont
  {Bartley}}, \bibinfo {author} {\bibfnamefont {Adriana~E.}\ \bibnamefont
  {Lita}}, \bibinfo {author} {\bibfnamefont {Sae~Woo}\ \bibnamefont {Nam}},
  \bibinfo {author} {\bibfnamefont {Thomas}\ \bibnamefont {Gerrits}}, \ and\
  \bibinfo {author} {\bibfnamefont {Christine}\ \bibnamefont {Silberhorn}},\
  }\bibfield  {title} {\enquote {\bibinfo {title} {{Single-Mode
  Parametric-Down-Conversion States with 50 Photons as a Source for Mesoscopic
  Quantum Optics}},}\ }\href {\doibase 10.1103/PhysRevLett.116.143601}
  {\bibfield  {journal} {\bibinfo  {journal} {Physical Review Letters}\
  }\textbf {\bibinfo {volume} {116}},\ \bibinfo {pages} {1--5} (\bibinfo {year}
  {2016})}\BibitemShut {NoStop}%
\bibitem [{\citenamefont {Bra{\'{n}}czyk}\ \emph {et~al.}(2010)\citenamefont
  {Bra{\'{n}}czyk}, \citenamefont {Ralph}, \citenamefont {Helwig},\ and\
  \citenamefont {Silberhorn}}]{Branczyk2010}%
  \BibitemOpen
  \bibfield  {author} {\bibinfo {author} {\bibfnamefont {Agata~M.}\
  \bibnamefont {Bra{\'{n}}czyk}}, \bibinfo {author} {\bibfnamefont {T.~C.}\
  \bibnamefont {Ralph}}, \bibinfo {author} {\bibfnamefont {Wolfram}\
  \bibnamefont {Helwig}}, \ and\ \bibinfo {author} {\bibfnamefont {Christine}\
  \bibnamefont {Silberhorn}},\ }\bibfield  {title} {\enquote {\bibinfo {title}
  {{Optimized generation of heralded Fock states using parametric
  down-conversion}},}\ }\href {\doibase 10.1088/1367-2630/12/6/063001}
  {\bibfield  {journal} {\bibinfo  {journal} {New Journal of Physics}\ }\textbf
  {\bibinfo {volume} {12}} (\bibinfo {year} {2010}),\
  10.1088/1367-2630/12/6/063001}\BibitemShut {NoStop}%
\bibitem [{\citenamefont {Christ}\ and\ \citenamefont
  {Silberhorn}(2012)}]{Christ2012}%
  \BibitemOpen
  \bibfield  {author} {\bibinfo {author} {\bibfnamefont {Andreas}\ \bibnamefont
  {Christ}}\ and\ \bibinfo {author} {\bibfnamefont {Christine}\ \bibnamefont
  {Silberhorn}},\ }\bibfield  {title} {\enquote {\bibinfo {title} {{Limits on
  the deterministic creation of pure single-photon states using parametric
  down-conversion}},}\ }\href {\doibase 10.1103/PhysRevA.85.023829} {\bibfield
  {journal} {\bibinfo  {journal} {Physical Review A}\ }\textbf {\bibinfo
  {volume} {85}},\ \bibinfo {pages} {023829} (\bibinfo {year}
  {2012})}\BibitemShut {NoStop}%
\bibitem [{\citenamefont {Eckstein}\ \emph {et~al.}(2011)\citenamefont
  {Eckstein}, \citenamefont {Christ}, \citenamefont {Mosley},\ and\
  \citenamefont {Silberhorn}}]{Eckstein2011}%
  \BibitemOpen
  \bibfield  {author} {\bibinfo {author} {\bibfnamefont {Andreas}\ \bibnamefont
  {Eckstein}}, \bibinfo {author} {\bibfnamefont {Andreas}\ \bibnamefont
  {Christ}}, \bibinfo {author} {\bibfnamefont {Peter~J.}\ \bibnamefont
  {Mosley}}, \ and\ \bibinfo {author} {\bibfnamefont {Christine}\ \bibnamefont
  {Silberhorn}},\ }\bibfield  {title} {\enquote {\bibinfo {title} {{Highly
  efficient single-pass source of pulsed single-mode twin beams of light}},}\
  }\href {\doibase 10.1103/PhysRevLett.106.013603} {\bibfield  {journal}
  {\bibinfo  {journal} {Physical Review Letters}\ }\textbf {\bibinfo {volume}
  {106}},\ \bibinfo {pages} {1--4} (\bibinfo {year} {2011})}\BibitemShut
  {NoStop}%
\bibitem [{\citenamefont {Christ}\ \emph {et~al.}(2013)\citenamefont {Christ},
  \citenamefont {Brecht}, \citenamefont {Mauerer},\ and\ \citenamefont
  {Silberhorn}}]{Christ2013}%
  \BibitemOpen
  \bibfield  {author} {\bibinfo {author} {\bibfnamefont {Andreas}\ \bibnamefont
  {Christ}}, \bibinfo {author} {\bibfnamefont {Benjamin}\ \bibnamefont
  {Brecht}}, \bibinfo {author} {\bibfnamefont {Wolfgang}\ \bibnamefont
  {Mauerer}}, \ and\ \bibinfo {author} {\bibfnamefont {Christine}\ \bibnamefont
  {Silberhorn}},\ }\bibfield  {title} {\enquote {\bibinfo {title} {{Theory of
  quantum frequency conversion and type-II parametric down-conversion in the
  high-gain regime}},}\ }\href {\doibase 10.1088/1367-2630/15/5/053038}
  {\bibfield  {journal} {\bibinfo  {journal} {New Journal of Physics}\ }\textbf
  {\bibinfo {volume} {15}} (\bibinfo {year} {2013}),\
  10.1088/1367-2630/15/5/053038}\BibitemShut {NoStop}%
\bibitem [{\citenamefont {Quesada}\ and\ \citenamefont
  {Sipe}(2015)}]{Quesada2015}%
  \BibitemOpen
  \bibfield  {author} {\bibinfo {author} {\bibfnamefont {Nicol{\'{a}}s}\
  \bibnamefont {Quesada}}\ and\ \bibinfo {author} {\bibfnamefont {J.~E.}\
  \bibnamefont {Sipe}},\ }\bibfield  {title} {\enquote {\bibinfo {title}
  {{Time-ordering effects in the generation of entangled photons using
  nonlinear optical processes}},}\ }\href {\doibase
  10.1103/PhysRevLett.114.093903} {\bibfield  {journal} {\bibinfo  {journal}
  {Physical Review Letters}\ }\textbf {\bibinfo {volume} {114}},\ \bibinfo
  {pages} {1--5} (\bibinfo {year} {2015})}\BibitemShut {NoStop}%
\bibitem [{\citenamefont {Mauerer}\ \emph {et~al.}(2009)\citenamefont
  {Mauerer}, \citenamefont {Avenhaus}, \citenamefont {Helwig},\ and\
  \citenamefont {Silberhorn}}]{Mauerer2009}%
  \BibitemOpen
  \bibfield  {author} {\bibinfo {author} {\bibfnamefont {Wolfgang}\
  \bibnamefont {Mauerer}}, \bibinfo {author} {\bibfnamefont {Malte}\
  \bibnamefont {Avenhaus}}, \bibinfo {author} {\bibfnamefont {Wolfram}\
  \bibnamefont {Helwig}}, \ and\ \bibinfo {author} {\bibfnamefont {Christine}\
  \bibnamefont {Silberhorn}},\ }\bibfield  {title} {\enquote {\bibinfo {title}
  {{How colors influence numbers: Photon statistics of parametric
  down-conversion}},}\ }\href {\doibase 10.1103/PhysRevA.80.053815} {\bibfield
  {journal} {\bibinfo  {journal} {Physical Review A - Atomic, Molecular, and
  Optical Physics}\ }\textbf {\bibinfo {volume} {80}},\ \bibinfo {pages} {1--5}
  (\bibinfo {year} {2009})}\BibitemShut {NoStop}%
\bibitem [{\citenamefont {Neuts}(1981)}]{Neuts1981}%
  \BibitemOpen
  \bibfield  {author} {\bibinfo {author} {\bibfnamefont {Marcel~F}\
  \bibnamefont {Neuts}},\ }\href@noop {} {\emph {\bibinfo {title}
  {{Matrix-geometric solutions in stochastic models: an algorithmic
  approach}}}}\ (\bibinfo  {publisher} {Courier Corporation},\ \bibinfo {year}
  {1981})\BibitemShut {NoStop}%
\bibitem [{\citenamefont {Bra{\'{n}}czyk}\ \emph {et~al.}(2011)\citenamefont
  {Bra{\'{n}}czyk}, \citenamefont {Fedrizzi}, \citenamefont {Stace},
  \citenamefont {Ralph},\ and\ \citenamefont {White}}]{Branczyk2011}%
  \BibitemOpen
  \bibfield  {author} {\bibinfo {author} {\bibfnamefont {Agata~M.}\
  \bibnamefont {Bra{\'{n}}czyk}}, \bibinfo {author} {\bibfnamefont
  {Alessandro}\ \bibnamefont {Fedrizzi}}, \bibinfo {author} {\bibfnamefont
  {Thomas~M.}\ \bibnamefont {Stace}}, \bibinfo {author} {\bibfnamefont
  {Tim~C.}\ \bibnamefont {Ralph}}, \ and\ \bibinfo {author} {\bibfnamefont
  {Andrew~G.}\ \bibnamefont {White}},\ }\bibfield  {title} {\enquote {\bibinfo
  {title} {{Engineered optical nonlinearity for quantum light sources}},}\
  }\href {\doibase 10.1364/OE.19.000055} {\bibfield  {journal} {\bibinfo
  {journal} {Optics Express}\ }\textbf {\bibinfo {volume} {19}},\ \bibinfo
  {pages} {55} (\bibinfo {year} {2011})}\BibitemShut {NoStop}%
\bibitem [{\citenamefont {Graffitti}\ \emph {et~al.}(2017)\citenamefont
  {Graffitti}, \citenamefont {Barrow}, \citenamefont {Proietti}, \citenamefont
  {Kundys},\ and\ \citenamefont {Fedrizzi}}]{Graffitti2017}%
  \BibitemOpen
  \bibfield  {author} {\bibinfo {author} {\bibfnamefont {Francesco}\
  \bibnamefont {Graffitti}}, \bibinfo {author} {\bibfnamefont {Peter}\
  \bibnamefont {Barrow}}, \bibinfo {author} {\bibfnamefont {Massimiliano}\
  \bibnamefont {Proietti}}, \bibinfo {author} {\bibfnamefont {Dmytro}\
  \bibnamefont {Kundys}}, \ and\ \bibinfo {author} {\bibfnamefont {Alessandro}\
  \bibnamefont {Fedrizzi}},\ }\bibfield  {title} {\enquote {\bibinfo {title}
  {{Independent high-purity photons created in domain-engineered crystals}},}\
  }\href {\doibase 10.1364/OPTICA.5.000514} {\bibfield  {journal} {\bibinfo
  {journal} {Optica}\ }\textbf {\bibinfo {volume} {5}},\ \bibinfo {pages}
  {1--4} (\bibinfo {year} {2017})}\BibitemShut {NoStop}%
\bibitem [{\citenamefont {U'Ren}\ \emph {et~al.}(2003)\citenamefont {U'Ren},
  \citenamefont {Banaszek},\ and\ \citenamefont {Walmsley}}]{URen2003}%
  \BibitemOpen
  \bibfield  {author} {\bibinfo {author} {\bibfnamefont {A.~B.}\ \bibnamefont
  {U'Ren}}, \bibinfo {author} {\bibfnamefont {K.}~\bibnamefont {Banaszek}}, \
  and\ \bibinfo {author} {\bibfnamefont {I.~A.}\ \bibnamefont {Walmsley}},\
  }\bibfield  {title} {\enquote {\bibinfo {title} {{Photon engineering for
  quantum information processing}},}\ }\href
  {http://arxiv.org/abs/quant-ph/0305192} {\ ,\ \bibinfo {pages} {1--23}
  (\bibinfo {year} {2003})},\ \Eprint {http://arxiv.org/abs/0305192}
  {arXiv:0305192 [quant-ph]} \BibitemShut {NoStop}%
\bibitem [{\citenamefont {Silberhorn}(2007)}]{Silberhorn2007}%
  \BibitemOpen
  \bibfield  {author} {\bibinfo {author} {\bibfnamefont {Christine}\
  \bibnamefont {Silberhorn}},\ }\bibfield  {title} {\enquote {\bibinfo {title}
  {{Detecting quantum light}},}\ }\href {\doibase 10.1080/00107510701662538}
  {\bibfield  {journal} {\bibinfo  {journal} {Contemporary Physics}\ }\textbf
  {\bibinfo {volume} {48}},\ \bibinfo {pages} {143--156} (\bibinfo {year}
  {2007})}\BibitemShut {NoStop}%
\bibitem [{\citenamefont {Sperling}\ \emph {et~al.}(2012)\citenamefont
  {Sperling}, \citenamefont {Vogel},\ and\ \citenamefont
  {Agarwal}}]{Sperling2012}%
  \BibitemOpen
  \bibfield  {author} {\bibinfo {author} {\bibfnamefont {J.}~\bibnamefont
  {Sperling}}, \bibinfo {author} {\bibfnamefont {W.}~\bibnamefont {Vogel}}, \
  and\ \bibinfo {author} {\bibfnamefont {G.~S.}\ \bibnamefont {Agarwal}},\
  }\bibfield  {title} {\enquote {\bibinfo {title} {{True photocounting
  statistics of multiple on-off detectors}},}\ }\href {\doibase
  10.1103/PhysRevA.85.023820} {\bibfield  {journal} {\bibinfo  {journal}
  {Physical Review A - Atomic, Molecular, and Optical Physics}\ }\textbf
  {\bibinfo {volume} {85}},\ \bibinfo {pages} {1--6} (\bibinfo {year}
  {2012})}\BibitemShut {NoStop}%
\bibitem [{\citenamefont {Jozsa}(1994)}]{Jozsa1994}%
  \BibitemOpen
  \bibfield  {author} {\bibinfo {author} {\bibfnamefont {Richard}\ \bibnamefont
  {Jozsa}},\ }\bibfield  {title} {\enquote {\bibinfo {title} {{Fidelity for
  mixed quantum states}},}\ }\href {\doibase 10.1080/09500349414552171}
  {\bibfield  {journal} {\bibinfo  {journal} {Journal of Modern Optics}\
  }\textbf {\bibinfo {volume} {41}},\ \bibinfo {pages} {2315--2323} (\bibinfo
  {year} {1994})}\BibitemShut {NoStop}%
\bibitem [{\citenamefont {Christ}\ \emph {et~al.}(2011)\citenamefont {Christ},
  \citenamefont {Laiho}, \citenamefont {Eckstein}, \citenamefont {Cassemiro},\
  and\ \citenamefont {Silberhorn}}]{Christ2011}%
  \BibitemOpen
  \bibfield  {author} {\bibinfo {author} {\bibfnamefont {Andreas}\ \bibnamefont
  {Christ}}, \bibinfo {author} {\bibfnamefont {Kaisa}\ \bibnamefont {Laiho}},
  \bibinfo {author} {\bibfnamefont {Andreas}\ \bibnamefont {Eckstein}},
  \bibinfo {author} {\bibfnamefont {Kat{\'{i}}scia~N.}\ \bibnamefont
  {Cassemiro}}, \ and\ \bibinfo {author} {\bibfnamefont {Christine}\
  \bibnamefont {Silberhorn}},\ }\bibfield  {title} {\enquote {\bibinfo {title}
  {{Probing multimode squeezing with correlation functions}},}\ }\href
  {\doibase 10.1088/1367-2630/13/3/033027} {\bibfield  {journal} {\bibinfo
  {journal} {New Journal of Physics}\ }\textbf {\bibinfo {volume} {13}}
  (\bibinfo {year} {2011}),\ 10.1088/1367-2630/13/3/033027}\BibitemShut
  {NoStop}%
\bibitem [{\citenamefont {Grice}\ \emph {et~al.}(2001)\citenamefont {Grice},
  \citenamefont {U'Ren},\ and\ \citenamefont {Walmsley}}]{Grice2001}%
  \BibitemOpen
  \bibfield  {author} {\bibinfo {author} {\bibfnamefont {W.~P.}\ \bibnamefont
  {Grice}}, \bibinfo {author} {\bibfnamefont {A.~B.}\ \bibnamefont {U'Ren}}, \
  and\ \bibinfo {author} {\bibfnamefont {I.~A.}\ \bibnamefont {Walmsley}},\
  }\bibfield  {title} {\enquote {\bibinfo {title} {{Eliminating frequency and
  space-time correlations in multiphoton states}},}\ }\href {\doibase
  10.1103/PhysRevA.64.063801} {\bibfield  {journal} {\bibinfo  {journal}
  {Physical Review A. Atomic, Molecular, and Optical Physics}\ }\textbf
  {\bibinfo {volume} {64}},\ \bibinfo {pages} {1--7} (\bibinfo {year}
  {2001})}\BibitemShut {NoStop}%
\bibitem [{\citenamefont {Avenhaus}\ \emph {et~al.}(2008)\citenamefont
  {Avenhaus}, \citenamefont {Coldenstrodt-Ronge}, \citenamefont {Laiho},
  \citenamefont {Mauerer}, \citenamefont {Walmsley},\ and\ \citenamefont
  {Silberhorn}}]{Avenhaus2008}%
  \BibitemOpen
  \bibfield  {author} {\bibinfo {author} {\bibfnamefont {M.}~\bibnamefont
  {Avenhaus}}, \bibinfo {author} {\bibfnamefont {H.~B.}\ \bibnamefont
  {Coldenstrodt-Ronge}}, \bibinfo {author} {\bibfnamefont {K.}~\bibnamefont
  {Laiho}}, \bibinfo {author} {\bibfnamefont {W.}~\bibnamefont {Mauerer}},
  \bibinfo {author} {\bibfnamefont {I.~A.}\ \bibnamefont {Walmsley}}, \ and\
  \bibinfo {author} {\bibfnamefont {C.}~\bibnamefont {Silberhorn}},\ }\bibfield
   {title} {\enquote {\bibinfo {title} {{Photon number statistics of multimode
  parametric down-conversion}},}\ }\href {\doibase
  10.1103/PhysRevLett.101.053601} {\bibfield  {journal} {\bibinfo  {journal}
  {Physical Review Letters}\ }\textbf {\bibinfo {volume} {101}},\ \bibinfo
  {pages} {1--4} (\bibinfo {year} {2008})}\BibitemShut {NoStop}%
\bibitem [{\citenamefont {Lita}\ \emph {et~al.}(2008)\citenamefont {Lita},
  \citenamefont {Miller},\ and\ \citenamefont {Nam}}]{Lita2008}%
  \BibitemOpen
  \bibfield  {author} {\bibinfo {author} {\bibfnamefont {Adriana~E.}\
  \bibnamefont {Lita}}, \bibinfo {author} {\bibfnamefont {Aaron~J.}\
  \bibnamefont {Miller}}, \ and\ \bibinfo {author} {\bibfnamefont {Sae~Woo}\
  \bibnamefont {Nam}},\ }\bibfield  {title} {\enquote {\bibinfo {title}
  {{Counting near-infrared single-photons with 95{\%} efficiency}},}\ }\href
  {\doibase 10.1364/OE.16.003032} {\bibfield  {journal} {\bibinfo  {journal}
  {Optics Express}\ }\textbf {\bibinfo {volume} {16}},\ \bibinfo {pages} {3032}
  (\bibinfo {year} {2008})}\BibitemShut {NoStop}%
\bibitem [{\citenamefont {Humphreys}\ \emph {et~al.}(2015)\citenamefont
  {Humphreys}, \citenamefont {Metcalf}, \citenamefont {Gerrits}, \citenamefont
  {Hiemstra}, \citenamefont {Lita}, \citenamefont {Nunn}, \citenamefont {Nam},
  \citenamefont {Datta}, \citenamefont {Kolthammer},\ and\ \citenamefont
  {Walmsley}}]{Humphreys2015}%
  \BibitemOpen
  \bibfield  {author} {\bibinfo {author} {\bibfnamefont {Peter~C.}\
  \bibnamefont {Humphreys}}, \bibinfo {author} {\bibfnamefont {Benjamin~J.}\
  \bibnamefont {Metcalf}}, \bibinfo {author} {\bibfnamefont {Thomas}\
  \bibnamefont {Gerrits}}, \bibinfo {author} {\bibfnamefont {Thomas}\
  \bibnamefont {Hiemstra}}, \bibinfo {author} {\bibfnamefont {Adriana~E.}\
  \bibnamefont {Lita}}, \bibinfo {author} {\bibfnamefont {Joshua}\ \bibnamefont
  {Nunn}}, \bibinfo {author} {\bibfnamefont {Sae~Woo}\ \bibnamefont {Nam}},
  \bibinfo {author} {\bibfnamefont {Animesh}\ \bibnamefont {Datta}}, \bibinfo
  {author} {\bibfnamefont {W.~Steven}\ \bibnamefont {Kolthammer}}, \ and\
  \bibinfo {author} {\bibfnamefont {Ian~A.}\ \bibnamefont {Walmsley}},\
  }\bibfield  {title} {\enquote {\bibinfo {title} {{Tomography of photon-number
  resolving continuous-output detectors}},}\ }\href {\doibase
  10.1088/1367-2630/17/10/103044} {\bibfield  {journal} {\bibinfo  {journal}
  {New Journal of Physics}\ }\textbf {\bibinfo {volume} {17}} (\bibinfo {year}
  {2015}),\ 10.1088/1367-2630/17/10/103044}\BibitemShut {NoStop}%
\bibitem [{\citenamefont {Burenkov}\ \emph {et~al.}(2017)\citenamefont
  {Burenkov}, \citenamefont {Sharma}, \citenamefont {Gerrits}, \citenamefont
  {Harder}, \citenamefont {Bartley}, \citenamefont {Silberhorn}, \citenamefont
  {Goldschmidt},\ and\ \citenamefont {Polyakov}}]{Burenkov2017}%
  \BibitemOpen
  \bibfield  {author} {\bibinfo {author} {\bibfnamefont {I~A}\ \bibnamefont
  {Burenkov}}, \bibinfo {author} {\bibfnamefont {A~K}\ \bibnamefont {Sharma}},
  \bibinfo {author} {\bibfnamefont {T}~\bibnamefont {Gerrits}}, \bibinfo
  {author} {\bibfnamefont {G}~\bibnamefont {Harder}}, \bibinfo {author}
  {\bibfnamefont {T~J}\ \bibnamefont {Bartley}}, \bibinfo {author}
  {\bibfnamefont {C}~\bibnamefont {Silberhorn}}, \bibinfo {author}
  {\bibfnamefont {E~A}\ \bibnamefont {Goldschmidt}}, \ and\ \bibinfo {author}
  {\bibfnamefont {S~V}\ \bibnamefont {Polyakov}},\ }\bibfield  {title}
  {\enquote {\bibinfo {title} {{Full statistical mode reconstruction of a light
  field via a photon-number-resolved measurement}},}\ }\href {\doibase
  10.1103/PhysRevA.95.053806} {\bibfield  {journal} {\bibinfo  {journal}
  {Physical Review A}\ }\textbf {\bibinfo {volume} {95}},\ \bibinfo {pages}
  {1--6} (\bibinfo {year} {2017})}\BibitemShut {NoStop}%
\bibitem [{\citenamefont {Dyakonov}\ \emph {et~al.}(2015)\citenamefont
  {Dyakonov}, \citenamefont {Sharapova}, \citenamefont {Iskhakov},\ and\
  \citenamefont {Leuchs}}]{Dyakonov2015}%
  \BibitemOpen
  \bibfield  {author} {\bibinfo {author} {\bibfnamefont {I.~V.}\ \bibnamefont
  {Dyakonov}}, \bibinfo {author} {\bibfnamefont {P.~R.}\ \bibnamefont
  {Sharapova}}, \bibinfo {author} {\bibfnamefont {T.~Sh}\ \bibnamefont
  {Iskhakov}}, \ and\ \bibinfo {author} {\bibfnamefont {G.}~\bibnamefont
  {Leuchs}},\ }\bibfield  {title} {\enquote {\bibinfo {title} {{Direct Schmidt
  number measurement of high-gain parametric down conversion}},}\ }\href
  {\doibase 10.1088/1612-2011/12/6/065202} {\bibfield  {journal} {\bibinfo
  {journal} {Laser Physics Letters}\ }\textbf {\bibinfo {volume} {12}}
  (\bibinfo {year} {2015}),\ 10.1088/1612-2011/12/6/065202}\BibitemShut
  {NoStop}%
\bibitem [{\citenamefont {Meyer-Scott}\ \emph {et~al.}(2017)\citenamefont
  {Meyer-Scott}, \citenamefont {Montaut}, \citenamefont {Tiedau}, \citenamefont
  {Sansoni}, \citenamefont {Herrmann}, \citenamefont {Bartley},\ and\
  \citenamefont {Silberhorn}}]{Meyer-Scott2017}%
  \BibitemOpen
  \bibfield  {author} {\bibinfo {author} {\bibfnamefont {Evan}\ \bibnamefont
  {Meyer-Scott}}, \bibinfo {author} {\bibfnamefont {Nicola}\ \bibnamefont
  {Montaut}}, \bibinfo {author} {\bibfnamefont {Johannes}\ \bibnamefont
  {Tiedau}}, \bibinfo {author} {\bibfnamefont {Linda}\ \bibnamefont {Sansoni}},
  \bibinfo {author} {\bibfnamefont {Harald}\ \bibnamefont {Herrmann}}, \bibinfo
  {author} {\bibfnamefont {Tim~J.}\ \bibnamefont {Bartley}}, \ and\ \bibinfo
  {author} {\bibfnamefont {Christine}\ \bibnamefont {Silberhorn}},\ }\bibfield
  {title} {\enquote {\bibinfo {title} {{Limits on the heralding efficiencies
  and spectral purities of spectrally filtered single photons from photon-pair
  sources}},}\ }\href {\doibase 10.1103/PhysRevA.95.061803} {\bibfield
  {journal} {\bibinfo  {journal} {Physical Review A}\ }\textbf {\bibinfo
  {volume} {95}},\ \bibinfo {pages} {1--8} (\bibinfo {year}
  {2017})}\BibitemShut {NoStop}%
\bibitem [{\citenamefont {Shalm}\ \emph {et~al.}(2015)\citenamefont {Shalm},
  \citenamefont {Meyer-Scott}, \citenamefont {Christensen}, \citenamefont
  {Bierhorst}, \citenamefont {Wayne}, \citenamefont {Stevens}, \citenamefont
  {Gerrits}, \citenamefont {Glancy}, \citenamefont {Hamel}, \citenamefont
  {Allman}, \citenamefont {Coakley}, \citenamefont {Dyer}, \citenamefont
  {Hodge}, \citenamefont {Lita}, \citenamefont {Verma}, \citenamefont
  {Lambrocco}, \citenamefont {Tortorici}, \citenamefont {Migdall},
  \citenamefont {Zhang}, \citenamefont {Kumor}, \citenamefont {Farr},
  \citenamefont {Marsili}, \citenamefont {Shaw}, \citenamefont {Stern},
  \citenamefont {Abell{\'{a}}n}, \citenamefont {Amaya}, \citenamefont
  {Pruneri}, \citenamefont {Jennewein}, \citenamefont {Mitchell}, \citenamefont
  {Kwiat}, \citenamefont {Bienfang}, \citenamefont {Mirin}, \citenamefont
  {Knill},\ and\ \citenamefont {Nam}}]{Shalm2015}%
  \BibitemOpen
  \bibfield  {author} {\bibinfo {author} {\bibfnamefont {Lynden~K.}\
  \bibnamefont {Shalm}}, \bibinfo {author} {\bibfnamefont {Evan}\ \bibnamefont
  {Meyer-Scott}}, \bibinfo {author} {\bibfnamefont {Bradley~G.}\ \bibnamefont
  {Christensen}}, \bibinfo {author} {\bibfnamefont {Peter}\ \bibnamefont
  {Bierhorst}}, \bibinfo {author} {\bibfnamefont {Michael~A.}\ \bibnamefont
  {Wayne}}, \bibinfo {author} {\bibfnamefont {Martin~J.}\ \bibnamefont
  {Stevens}}, \bibinfo {author} {\bibfnamefont {Thomas}\ \bibnamefont
  {Gerrits}}, \bibinfo {author} {\bibfnamefont {Scott}\ \bibnamefont {Glancy}},
  \bibinfo {author} {\bibfnamefont {Deny~R.}\ \bibnamefont {Hamel}}, \bibinfo
  {author} {\bibfnamefont {Michael~S.}\ \bibnamefont {Allman}}, \bibinfo
  {author} {\bibfnamefont {Kevin~J.}\ \bibnamefont {Coakley}}, \bibinfo
  {author} {\bibfnamefont {Shellee~D.}\ \bibnamefont {Dyer}}, \bibinfo {author}
  {\bibfnamefont {Carson}\ \bibnamefont {Hodge}}, \bibinfo {author}
  {\bibfnamefont {Adriana~E.}\ \bibnamefont {Lita}}, \bibinfo {author}
  {\bibfnamefont {Varun~B.}\ \bibnamefont {Verma}}, \bibinfo {author}
  {\bibfnamefont {Camilla}\ \bibnamefont {Lambrocco}}, \bibinfo {author}
  {\bibfnamefont {Edward}\ \bibnamefont {Tortorici}}, \bibinfo {author}
  {\bibfnamefont {Alan~L.}\ \bibnamefont {Migdall}}, \bibinfo {author}
  {\bibfnamefont {Yanbao}\ \bibnamefont {Zhang}}, \bibinfo {author}
  {\bibfnamefont {Daniel~R.}\ \bibnamefont {Kumor}}, \bibinfo {author}
  {\bibfnamefont {William~H.}\ \bibnamefont {Farr}}, \bibinfo {author}
  {\bibfnamefont {Francesco}\ \bibnamefont {Marsili}}, \bibinfo {author}
  {\bibfnamefont {Matthew~D.}\ \bibnamefont {Shaw}}, \bibinfo {author}
  {\bibfnamefont {Jeffrey~A.}\ \bibnamefont {Stern}}, \bibinfo {author}
  {\bibfnamefont {Carlos}\ \bibnamefont {Abell{\'{a}}n}}, \bibinfo {author}
  {\bibfnamefont {Waldimar}\ \bibnamefont {Amaya}}, \bibinfo {author}
  {\bibfnamefont {Valerio}\ \bibnamefont {Pruneri}}, \bibinfo {author}
  {\bibfnamefont {Thomas}\ \bibnamefont {Jennewein}}, \bibinfo {author}
  {\bibfnamefont {Morgan~W.}\ \bibnamefont {Mitchell}}, \bibinfo {author}
  {\bibfnamefont {Paul~G.}\ \bibnamefont {Kwiat}}, \bibinfo {author}
  {\bibfnamefont {Joshua~C.}\ \bibnamefont {Bienfang}}, \bibinfo {author}
  {\bibfnamefont {Richard~P.}\ \bibnamefont {Mirin}}, \bibinfo {author}
  {\bibfnamefont {Emanuel}\ \bibnamefont {Knill}}, \ and\ \bibinfo {author}
  {\bibfnamefont {Sae~Woo}\ \bibnamefont {Nam}},\ }\bibfield  {title} {\enquote
  {\bibinfo {title} {{Strong Loophole-Free Test of Local Realism}},}\ }\href
  {\doibase 10.1103/PhysRevLett.115.250402} {\bibfield  {journal} {\bibinfo
  {journal} {Physical Review Letters}\ }\textbf {\bibinfo {volume} {115}},\
  \bibinfo {pages} {1--10} (\bibinfo {year} {2015})}\BibitemShut {NoStop}%
\end{thebibliography}%


%

\clearpage
\newpage

Supplemental material:\\

\paragraph*{Convolution of spectral modes with equal strength:\ --} 
Here we will consider spectral modes with the same vacuum probability $q$ to derive the heralding probability. Every mode is a geometric distribution:
\begin{equation}
p_n(q) = q(1-q)^n~,
\label{eq:02}
\end{equation}
with the vacuum probability $q=1-|\Lambda|^2$.
The convolution of two modes $p_{n,2}$ is given by
\begin{equation}
p_{n,2}(q) = \sum_{i=0}^n p_i(q)\cdot p_{n-i}(q)~.
\label{eq:03}
\end{equation}
This can be extended for $K$ spectral modes if the new distribution is convoluted recursively with a geometric distribution. The resulting distribution after $K$ steps is known as a negative binomial distribution
\begin{equation}
p_{n,K}(q) = \binom{n+K-1}{n}(1-q)^n q^K~.
\label{eq:negbin}
\end{equation}
In the limit of infinite modes the distributions becomes Possonian and the maximal probability to herald an $n$-photon Fock state is  
\begin{equation}
p_\text{max}(n) = \frac{e^{-n}n^n}{n!}~.
\label{eq:08}
\end{equation}
Likewise we can investigate the maximal fidelity of the heralded state towards an $n$-photon state in the first Schmidt mode and vacuum in all the other modes  ($\mathcal{F}(\rho,\sigma) = (tr\sqrt{\sqrt{\rho}\sigma\sqrt{\rho}})^2$ ). 
For spectral modes that have the same vacuum probability an analytic solution can be found: 
\begin{equation}
\mathcal{F}_\text{max} = \frac{(K-1)!n!}{(K+n-1)!}~.
\label{eq:09}
\end{equation}
For the more general case where the vacuum probabilities $q_k$ are strictly decreasing,  one can find at least an expression for the heralding probability 
\begin{equation}
p(n) = \prod_{i=1}^{K_\text{max}} q_i\cdot\sum_{j=1}^{K_\text{max}} \frac{(-1)^j(1-q_j)^{n+K_\text{max}-1}}{\prod_{m \neq j} |q_j-q_m|}~.
\label{eq:08}
\end{equation}

\begin{figure}[h!]
   \includegraphics[width=\linewidth]{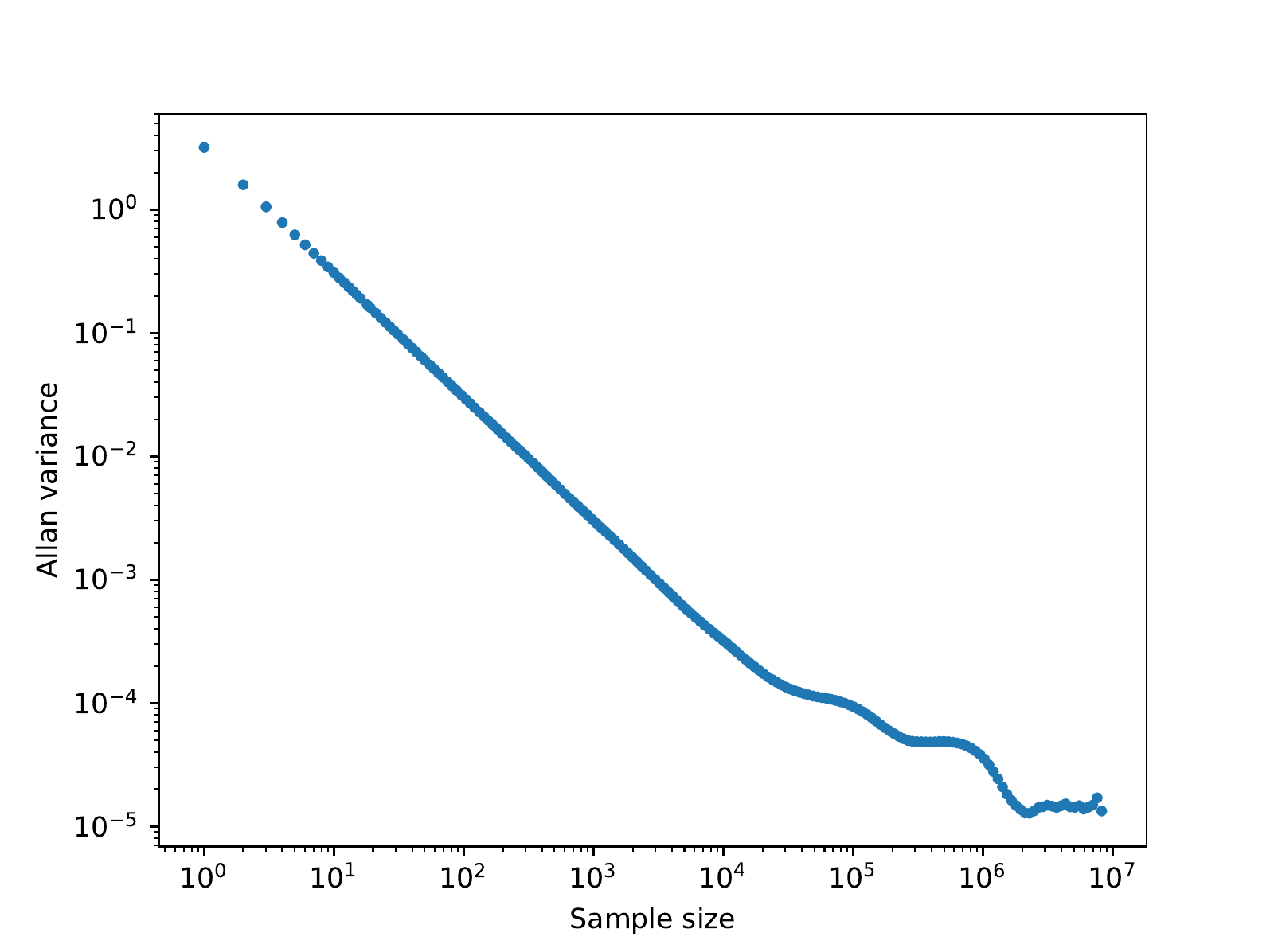}
\caption{Allan variance of the measured photon numbers. The plot shows that there is no measurable drift (e.g. in waveguide coupling) over one measurement run as the Allan variance is decreasing with sample size.}
\label{fig:allan_var}
\end{figure}

\begin{figure}[h!]
   \includegraphics[width=\linewidth]{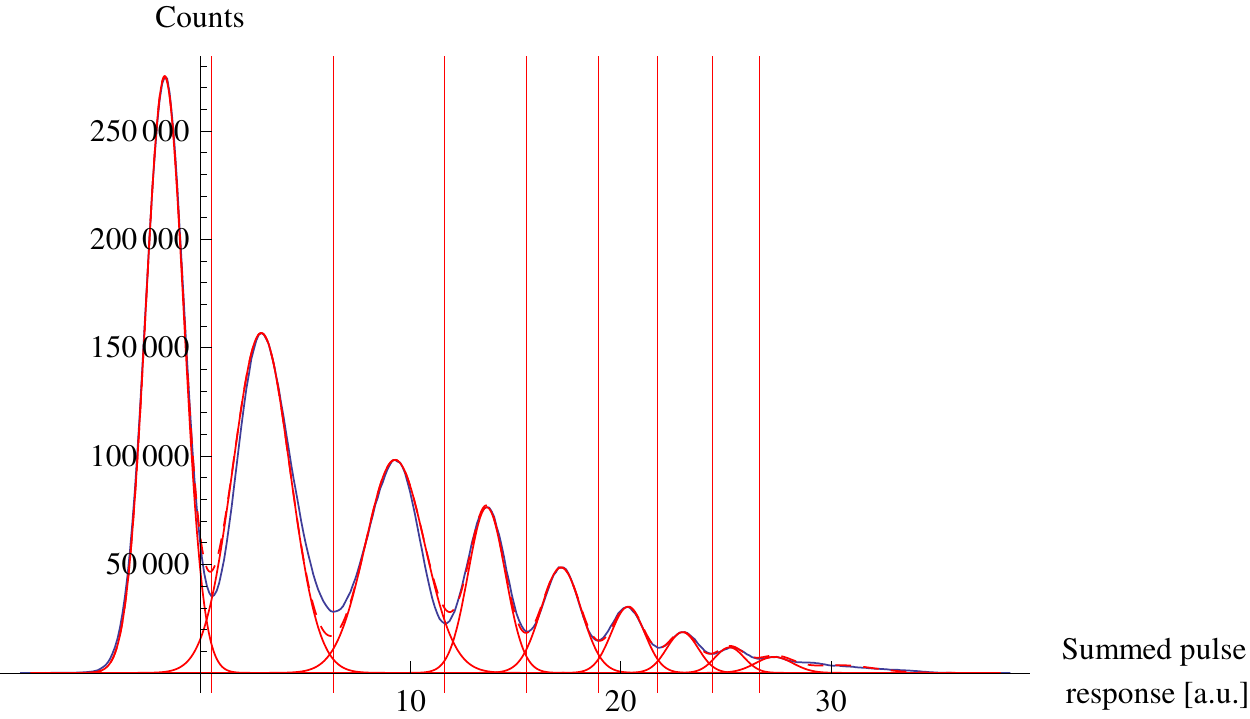}
\caption{Histogram showing summed pulse response from a TES (blue curve) multiplied by an average pulse response. First peak corresponds to a vacuum event, second peak to a detected single photon etc. The sum of multiple Gaussian functions where fitted to the data (dashed red curve). The individual Gaussian functions (red curves) are used to calculate the error bars.}
\label{fig:tesres}
\end{figure}

\paragraph*{Error bars\ --}
The error bars shown in Fig.~\ref{fig:exp} can be decomposed into three different error sources. A) Statistical errors due to finite measurement time. This error source is very small for most data points as the sample size is large. For the data points were less than 10 000 data events were present the statistical uncertainties were calculated based on estimating the success probability of a Bernoulli process. 
B) Time dependent fluctuations in the setup. This effect could for example be caused by thermal fluctuations in the lab resulting in coupling drifts. We analyzed this effect using the Allan-Variance (Fig.~\ref{fig:allan_var}). However the data within each of the four measurements only showed negligible signs of drifts.  C) Converting the continuous output function from TESs into photon numbers. For this experiment this conversion is the main error source. This conversion is a known problem which is analyzed in detail in \cite{Humphreys2015}. Here we use a slightly different approach of assuming Gaussian response functions for the individual photon numbers and fitting them to the data. Fig.~\ref{fig:tesres} shows a histogram of the voltage signal from the TES. The voltage signal for an event here was summed and multiplied by an average voltage response curve. Each peak corresponds to a certain photon number. Vertical lines show the acceptance windows for the different photon numbers. The fitted Gaussians were used to calculate the probability that the response of a certain photon number was not inside the acceptance window and the probability that a click inside the acceptance window was caused by a different photon number event. These two probabilities cause asymmetric error bars  as shown in Fig.~\ref{fig:exp}. 
The error could be reduced by using narrower acceptance windows as suggested in \cite{Humphreys2015}. However, this method is not used in the context of this paper as this would decrease the heralding probability.  \\

\paragraph*{Schmidt decomposition\ --}
As shown in \cite{Christ2011} we perform a Schmidt decomposition on the spectral correlation function $f(\omega_\text{s}, \omega_\text{i})$ of the PDC Hamiltonian 
\begin{equation}
H_\text{PDC} = A \int d\omega_\text{s} \int d\omega_\text{i} f(\omega_\text{s}, \omega_\text{i}) a^\dag_\text{s}(\omega_\text{s}) a^\dag_\text{i}(\omega_\text{i}) + h.c.~.
\label{eq:HPDC}
\end{equation}
Here A is a constant characterizing the interaction strength. We want to point out that squeezing value and Schmidt decomposition are independent as the Schmidt decomposition is performed on the spectral correlation function $f$ of the operators. 
However as shown in \cite{Dyakonov2015} the relative number of photons per mode is changing with pump power. This means that at high gain most of the photons will be produced in the dominant Schmidt mode. Therefore sensitive measures to this value like the $g^{(2)}$ should be used carefully in the high gain regime \cite{Christ2011}. We want to stress here again that changing the pump intensity does not change the Schmidt number.  
A Schmidt decomposition requires two sets of orthonormal functions and therefore it is not possible to rescale the Schmidt coefficients in the high gain regime. 
 
It should be noted that the state fidelity is very sensitive to the Schmidt number. This is different from other applications where for example a photon is subtracted. In that case increasing the gain can help to compensate the multimodeness because the relative number of photons per mode is changing with gain. In the high gain regime the dominant mode is amplified most and therefore increasing the probability to subtract a photon from this mode. The important point is that heralding higher order Fock states is different from this photon subtraction scenario. Multimodeness will result in a lower state fidelity as here a specific photon number $n$ is fixed by the heralding process. The ratio of the generation probability (for a fixed $n$) of the first Schmidt mode and the sum of all other modes is strictly decreasing with increasing pump intensity. 
This means that choosing the appropriate phase matching and pump bandwidth of the SPDC process (typically known as source engineering) is very important for heralding Fock states. Spectral filtering is also not an option as this will increase losses in the system \cite{Meyer-Scott2017}. \\

\paragraph*{Fidelity\ --}
In order to quantify the quality of the heralded state, the fidelity to a desired Fock state $n$ in a given spectral mode $k$ will be calculated. This spectral mode is chosen to be the largest mode in the Schmidt decomposition of the state ($k=1$). We will use state fidelity from now on for this value. At first we will consider no losses for the signal arm, whose effects are then added in the second step. 

Fidelity without losses for the signal:\\
The state fidelity can be calculated from the probability to generate $n$ photons in the dominant mode $p_\text{n,k=1}$ times the probability to have vacuum in all the other modes $p_\text{0,k$>$1}$, divided by the heralding probability $p_\text{n}$ for this photon number. If photons were produced in other spectral modes they would also be present in the heralded state as no losses are present there. Losses in the idler arm will influence the heralding probability $p_\text{n}$.

Fidelity with losses for the signal:\\
If losses in the signal arm are present more cases need to be considered. 
For example we can investigate the case where 1 photon was incident on the heralding detector. Two possibilities can now lead to the desired single photon state in the correct spectral mode. A: one photon from the correct spectral mode caused the heralding event and one photon from this mode also was transmitted in the signal arm while no photon in the other modes where produced or they were lost. B: no photons from the first mode reach the heralding detector but one photon from this mode is transmitted through the signal arm while one photon from one of the other modes is causing the heralding event and no photons from the other modes are transmitted through the signal arm.\\ 
Generalizing this example yields 
\begin{equation}
\mathcal{F}(\ket{n}\!\bra{n}_\text{k=1},\rho) =  \frac{1}{p_\text{n}}\sum_{i=0}^n p_\text{k=1}(n,n-i)p_\text{k$>$1}(0,i)~,
\label{eq:fid_gen}
\end{equation}
with a heralding probability $p_\text{n}$ from all modes, a probability $p_\text{k=1}(n,n-i)$ of $n$ photons in the signal arm and $n-i$ photons in the idler arm of the first spectral mode and a probability $p_\text{k$>$1}(0,i)$ of zero photons in the signal arm and $i$ photons in the idler arm from all other spectral modes. All probabilities have to be calculated after applying the loss operations. It should be noted that losses in the idler arm can actually increase the state fidelity for a fixed amount of loss in the signal arm. This effect comes from the fact that the desired photon number probability for the signal $p_\text{k=1}(n)$ can be increased from higher order photon number contributions and losses contributing to this probability. In specific cases it can be beneficial for the state fidelity to herald on a different photon number $j$ than the desired fock state $n\neq j$. These cases are not considered here as they only occur for high photon numbers or low transmission values. 

\begin{figure}[h!]
\includegraphics[width=\linewidth]{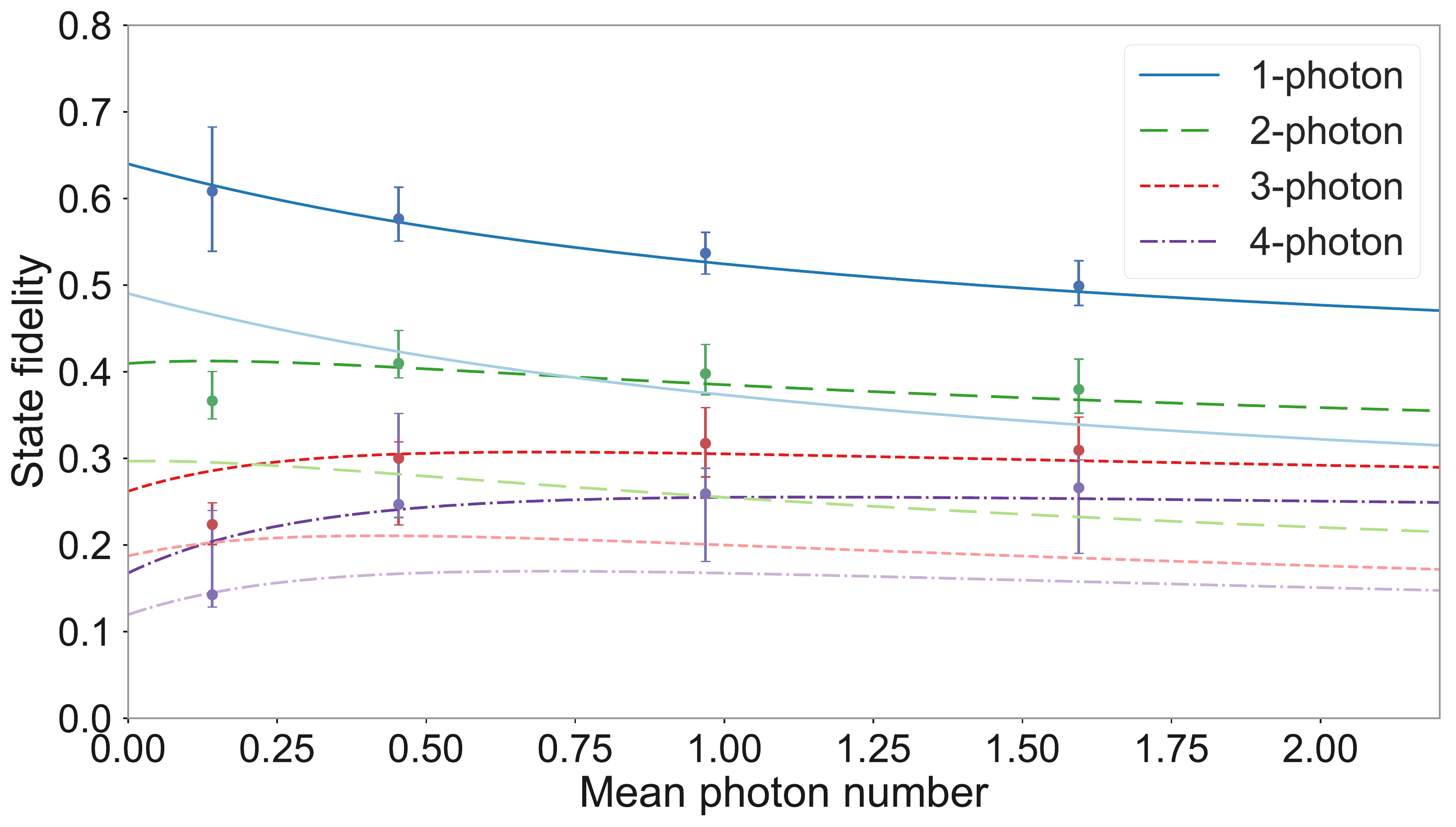}
\caption{Experimental data (points) the state fidelity (b) in dependence of the generation probability. Four different pump intensities were investigated. Colored lines show the theoretical values with $K=1.61$, $\eta_i=0.59$ and $\eta_s=0.64$ as the only free parameters. 
In (b) strong colors and the measured data points, show the case where spectral multimodeness is not considered. Light colors include spectral multimodeness. See text for further details.}
\label{fig:exp2}
\end{figure}

As shown in the main text the Schmidt number and detection efficiency can be extracted from the measured data. Based on these values the single mode fidelity of the heralded state towards the desired Fock state in the strongest spectral mode can be extracted. This plotted in light colors (Fig.~\ref{fig:exp2}).  \\

\paragraph*{Realistic limits for generating Fock states\ --}
In order to calculate the highest Fock state $n$ that can be generated with SPDC, we assume optimistic values for an experiment. We choose a fidelity of $90\%$ with respect to the desired Fock state as a threshold for acceptable quality. We will not consider losses in the signal arm and assume perfect emission into a single Schmidt mode.  \\
Heralding efficiency: We will take a heralding efficiency of $\eta = 90\%$. This is above the highest value that has been shown in the literature (see e.g. \cite{Shalm2015}), but may be achievable in principle. 
Generation probability: In order to acquire enough statistics in a reasonable time we aim for 0.1 event/s. The maximal repetition rate of the experiment is limited by the laser system as well as the detection system. Typical laser systems used for these experiments are Ti:Sapphire laser, which offer repetition rates around $10^8$ pulses/s, as well as high energies in transform-limited pulses. We will chose this value as an optimistic value for the repetition rate although the detection part of the experiment may be slower. 
With these restrictions it can be shown that only Fock states up to $n=9$ can be realized at 0.1~event/s.\\

\paragraph*{Spectral mode changes under losses\ --}
A thermal state will stay thermal under losses. In addition we can describe a thermal state completely by analyzing the vacuum probability $q$ (cf. \eqref{eq:02}). This means that we can find an expression for a new thermal state with vacuum probability $q'$ after a transmission value of $\eta$
\begin{equation}
\begin{aligned}
q' &= \sum^\infty_{i=0} \binom{i}{0}(1-\eta)^i\eta^0 \cdot q(1-q)^i\\
&= \frac{q}{q+\eta-q\eta}~.
\label{sup01}
\end{aligned}
\end{equation}
Expressing the vacuum probability by the squeezing parameter yields 
\begin{equation}
r' = \arctanh\left(\sqrt{\frac{\eta \tanh(r)^2}{1+(\eta-1)\tanh(r)^2}}\right)~.
\label{sup02}
\end{equation}
\begin{figure}[h!]
   \includegraphics[width=\linewidth]{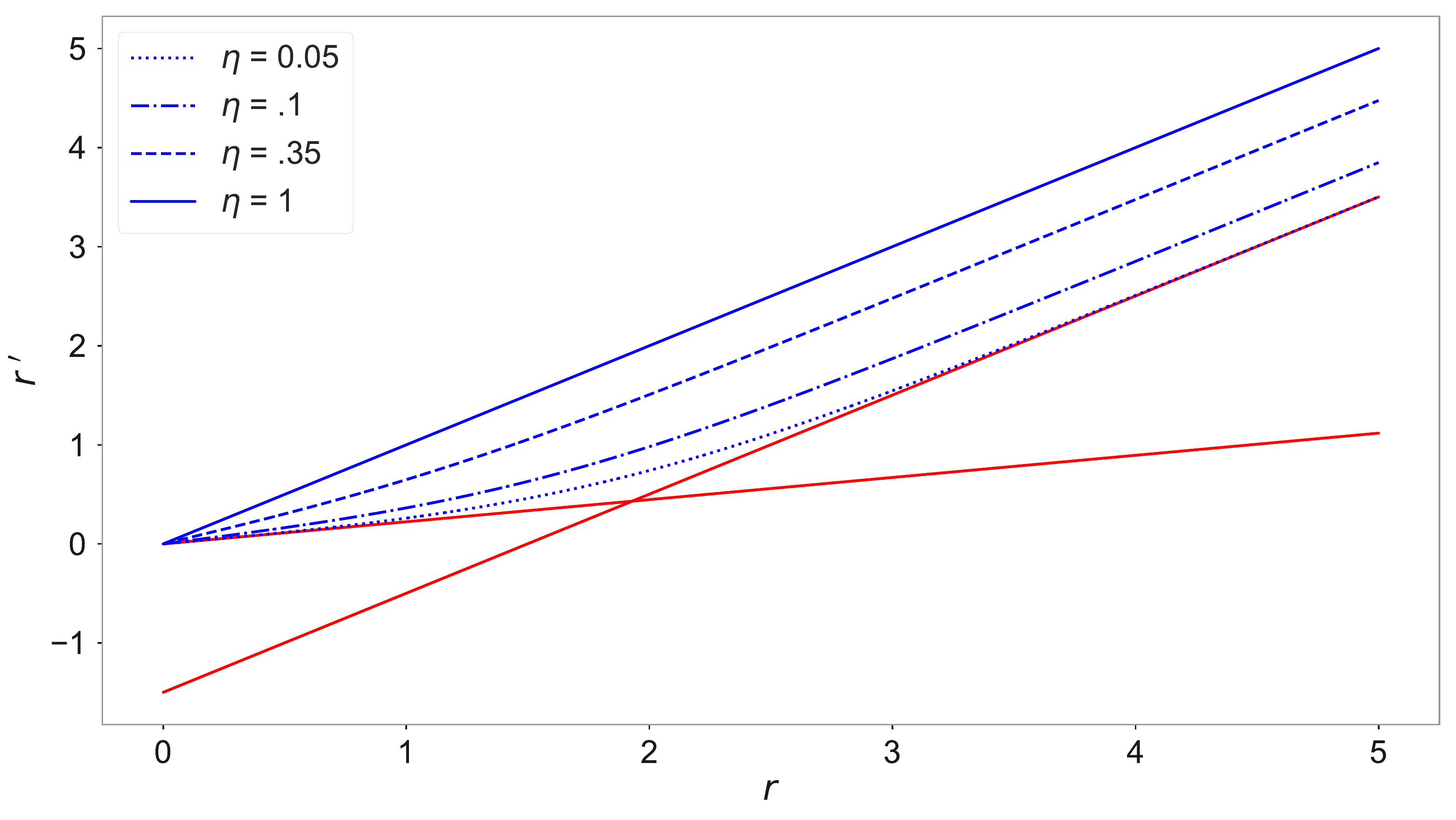}
\caption{Visualizing the relationship between a squeezing parameter of a thermal state before and after loss. Red lines illustrate the two linear regions with a slope of $\sqrt{\eta}$ and 1. }
\label{fig:sup01}
\end{figure}
Fig.~\ref{fig:sup01} visualizes Eq.~\ref{sup02}. Two different linear regions can be identified. Region A has the slope of $\sqrt{\eta}$ region B has a slope of 1.
The ratio between multiple spectral modes can only be preserved under losses if low squeezing parameters are present (all squeezing parameters in region A) such that $r'$ is linear in $r$. In general their ratio will change  although all individual thermal states stay thermal under losses. This is an important result as time ordering effects (which are not considered here) also predict the spectral changes in this squeezing regime. \\

\end{document}